\documentclass[twocolumn]{autart}    

\usepackage{amsmath,amssymb}
\usepackage[shortlabels]{enumitem}
\usepackage{xcolor}
\usepackage{subcaption}

\usepackage[title]{appendix}

\usepackage{graphicx}          

\usepackage[longnamesfirst]{natbib}
\setcitestyle{authoryear,aysep={,}}
\AtBeginDocument{}

\begin{document}

\begin{frontmatter}

\title{Signaling Games in Multiple Dimensions: Geometric Properties of Equilibrium Solutions\thanksref{footnoteinfo}} 

\thanks[footnoteinfo]{The material in this paper was partially presented at the 2021 International Symposium on Modeling and Optimization in Mobile, Ad hoc, and Wireless Networks (WiOpt), October 18–21, 2021, Philadelphia, PA, USA \citep*{KazikliWiOpt21}.}

\author[Tobb]{Ertan Kaz{\i}kl{\i}}\ead{ekazikli@etu.edu.tr},
\author[Bilkent]{Sinan Gezici}\ead{gezici@ee.bilkent.edu.tr},
\author[Queens]{Serdar Y\"uksel}\ead{yuksel@mast.queensu.ca}

\address[Tobb]{Electrical and Electronics Engineering, TOBB University of Economics and Technology, Ankara, Turkey}
\address[Bilkent]{Electrical and Electronics Engineering, Bilkent University, Ankara, Turkey}
\address[Queens]{Mathematics and Statistics, Queen's University, Kingston, Ontario, Canada}


\begin{abstract}
Signaling game problems investigate communication scenarios where encoder(s) and decoder(s) have misaligned objectives due to the fact that they either employ different cost functions or have inconsistent priors. This problem has been studied in the literature for scalar sources under various setups. In this paper, we consider multi-dimensional sources under quadratic criteria in the presence of a bias leading to a mismatch in the criteria, where we show that the generalization from the scalar setup is more than technical. We show that the Nash equilibrium solutions lead to structural richness due to the subtle geometric analysis the problem entails, with consequences in both system design, the presence of linear Nash equilibria, and an information theoretic problem formulation. We first provide a set of geometric conditions that must be satisfied in equilibrium considering any multi-dimensional source. Then, we consider independent and identically distributed sources and characterize necessary and sufficient conditions under which an informative linear Nash equilibrium exists. These conditions involve the bias vector that leads to misaligned costs. Depending on certain conditions related to the bias vector, the existence of linear Nash equilibria requires sources with a Gaussian or a symmetric density. Moreover, in the case of Gaussian sources, our results have a rate-distortion theoretic implication that achievable rates and distortions in the considered game theoretic setup can be obtained from its team theoretic counterpart.
\end{abstract}

\end{frontmatter}

\section{Introduction}\label{sec:intro}

In a team theoretic setup where the decision makers share a common goal, the decision makers do not wish to hide information to improve the performance since revealing more information does not lead to a degradation of system performance. Therefore, in such setups, if there is no constraint on messages to transmit between the decision makers, such as a power constraint or a limited bandwidth requirement, a decision maker can always reveal more information without causing any performance loss. On the other hand, in a game theoretic (strategic) setup involving decision makers with misaligned goals, revealing more information may hurt some or even all of the decision makers \citep*{Bassan03}. Hence, a decision maker in a strategic setting needs to take misaligned goals into account while designing what information to reveal to another decision maker. We may consider two main themes which lead to misaligned objectives for the decision makers. In the first theme, the decision makers employ different cost functions, e.g., a decision maker wishes to mislead another decision maker, see, e.g., \citep*{SaritasTAC2017}, \citep*{AkyolProcIEEE2017} and \citep*{TreustJET2019}. The second theme is concerned with the case when the decision makers have subjective beliefs regarding prior probability distributions of unknown parameters. This subjectivity leads to misaligned objectives for the decision makers even though they employ the same cost function, see, e.g., \citep*{BasarTAC85}, \citep*{MismatchArxiv} and \citep*{SaritasTSP2019}. These both lead to a game theoretic setup where a suitable equilibrium concept, such as the Nash equilibrium and the Stackelberg equilibrium, is to be used to analyze the system. These problems fall into the general class of signaling game problems that investigates communication scenarios between decision makers with misaligned objectives. In this context, Crawford and Sobel, in their seminal paper \citep*{CrawfordSobel}, introduce a signaling game problem where a biased encoder wishes to convey a scalar source to a decoder, and a message transmission does not induce a cost for the encoder. This problem is also referred to as {\it cheap talk}, which emphasizes that communication is costless. Crawford and Sobel show that under certain technical conditions regarding cost functions, the encoder must hide information at a Nash equilibrium by employing quantization policies, which holds even though there is no restriction on communication. In an equilibrium with a quantization policy, referred to as quantized or partition equilibrium, the encoder partitions the observation space into intervals and reveals the interval that contains the encoder's observation. Crawford and Sobel's result implies that at a Nash equilibrium, the encoder cannot convey its private information completely by employing a linear encoding policy (i.e., transmitting a scaled version of its observation to the decoder). This is a striking example where providing more information to the decoder by employing a linear encoder instead of a quantized encoder breaks the equilibrium in a game theoretic setup. In this manuscript, we study multi-dimensional sources under quadratic criteria for the cheap talk setup of Crawford and Sobel and investigate the properties of Nash equilibrium solutions.

Our work investigates communication scenarios between a biased encoder and a decoder, which leads to a signaling game problem. We may encounter biased decision makers in various applications. For instance, in control applications, an adversary may wish to inject a bias into a control system in order to deteriorate the system performance \citep{TeixeiraAutomatica15}. In smart grid applications, a strategic consumer or electricity producer in a microgrid system may wish to give false or biased measurement reports to another decision maker for its own benefit \citep{CSLloydMaxICASSP14}. As another application, strategic users in a cellular network may wish to misreport their channel conditions to the base station for their own benefit \citep{OpportunisticIT12}. Moreover, interactions between attackers and defenders in control applications may be modeled as a cheap talk problem \citep{ZuxingACC2020,SaritasACC20}. For applications of signaling games and cheap talk in fields such as economics, finance, biology, and political science, the reader is referred to \citep{Sobel2020}.

\begin{figure}
\centering
\includegraphics[width=0.5\linewidth]{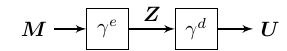}
\caption{Communication setting.}
\label{fig:block}
\end{figure} 

\subsection{Preliminaries}

We consider the following multi-dimensional signaling game problem where an encoder and a decoder communicate. This, in particular, corresponds to a multi-dimensional cheap talk problem where cheap talk refers to the fact that a message transmission does not induce a cost for the encoder. The encoder observes the value of an $n$-dimensional random vector $\boldsymbol{M}=[M_1,\dots,M_n]^T$ where $M_1,\dots,M_n$ are $\mathbb{M}$-valued random variables. The encoder conveys a message $\boldsymbol{Z}=[Z_1,\dots,Z_n]^T$ via an encoding policy $\gamma^e(\cdot)$, i.e., $\boldsymbol{Z}=\gamma^e(\boldsymbol{M})$, where $Z_1,\dots,Z_n$ are $\mathbb{Z}$-valued random variables. The decoder directly observes $\boldsymbol{Z}$ and takes an action $\boldsymbol{U}=[U_1,\dots,U_n]^T$ via a decoding policy $\gamma^d(\cdot)$, i.e., $\boldsymbol{U}=\gamma^d(\boldsymbol{Z})$, where $U_1,\dots,U_n$ are $\mathbb{M}$-valued random variables. In this paper, we consider real valued random variables, i.e., $\mathbb{M}=\mathbb{Z}=\mathbb{R}$ where $\mathbb{R}$ denotes the set of real numbers. The aim of the encoder is to minimize $J^e(\gamma^e,\gamma^d)=\mathbb{E}[c^e(\boldsymbol{M},\boldsymbol{U})]$ where\footnote{We adopt the convention that random variables are denoted by uppercase letters with their realizations denoted by the corresponding lowercase letters.}
\begin{align}
c^e(\boldsymbol{m},\boldsymbol{u}) 
= \sum_{i=1}^n (m_i-u_i-b_i)^2 
= \lVert \boldsymbol{m}-\boldsymbol{u}-\boldsymbol{b}\rVert^2.
\label{eq:encCost}
\end{align}
In \eqref{eq:encCost}, $\boldsymbol{b}$ denotes a deterministic bias vector which is common knowledge among the players and quantifies the degree of misalignment between the objective functions of the encoder and decoder. In other words, the encoder wishes to make biased reports regarding its observations possibly with different biases for different components. On the other hand, the decoder wishes to estimate the random source vector as accurately as possible; thus, its objective function does not include a bias vector. In particular, the aim of the decoder is to minimize $J^d(\gamma^e,\gamma^d)=\mathbb{E}[c^d(\boldsymbol{M},\boldsymbol{U})]$ where
\begin{align}
c^d(\boldsymbol{m},\boldsymbol{u}) 
= \sum_{i=1}^n (m_i-u_i)^2 
= \lVert \boldsymbol{m}-\boldsymbol{u}\rVert^2.
\label{eq:decCost}
\end{align}
The communication scenario is depicted in Fig.\ref{fig:block}. Our aim is to characterize the Nash equilibrium where the decision makers announce their policies at the same time. At a Nash equilibrium, none of the players wishes to unilaterally deviate from their current strategies as their cost cannot get better by doing so. In particular, a set of policies $\gamma^{*,e}$ and $\gamma^{*,d}$ forms a Nash equilibrium (e.g., \cite*{TBasarBook}) if 
\begin{align}
\begin{split}
J^e(\gamma^{*,e}, \gamma^{*,d}) &\leq J^e(\gamma^{e}, \gamma^{*,d})  \text{ for all } \gamma^e \in \Gamma^e ,\\
J^d(\gamma^{*,e}, \gamma^{*,d}) &\leq J^d(\gamma^{*,e}, \gamma^{d})  \text{ for all }\gamma^d \in \Gamma^d ,
\label{eq:nashEquilibrium}
\end{split}
\end{align}
where $\Gamma^e$ and $\Gamma^d$ are the sets of all deterministic (and Borel measurable) functions from $\mathbb{M}^n$ to $\mathbb{Z}^n$ and from $\mathbb{Z}^n$ to $\mathbb{M}^n$, respectively.

\begin{rem}
{\normalfont
Under the Nash equilibrium concept, both players announce their policies at the same time. By considering the Nash equilibrium concept, we essentially investigate a non-cooperative communication setup in terms of policy announcements in the sense that no player discloses its policy before the other player. This means that no player commits to a certain announced policy \emph{a priori}. This equilibrium concept is appropriate, for instance, when the players do not have access to policy announcements of each other or when they do not trust an announced policy by the other player. In contrast, one can also consider the Stackelberg setup (see, e.g., \citep{TBasarBook} for a definition) where the encoder announces its policy and commits to this policy, and the decoder chooses its policy given the encoder's announcement. We may view the Stackelberg setup as a cooperative communication setup as there is a policy announcement by the encoder. In fact, for the scalar or multi-dimensional cheap talk setup, the Stackelberg equilibrium solution leads to full revelation where the encoder discloses the source completely \citep[Theorem~3.3]{SaritasTAC2017}. In contrast, there does not exist a Nash equilibrium with full information revelation in general. In other words, the encoder must hide information partially (or even completely in certain cases, see, e.g., \citep[Theorem~3]{NumberOfBinsArxiv} for the scalar case) in the non-cooperative communication setup whereas it does not hide any information in the cooperative communication setup. 
}
\end{rem}

\begin{assum}\label{assumption}
Considering each component $M_i$ of the source random vector $\boldsymbol{M}$, every non-empty open set on its support has a positive measure. 
\end{assum}
The following is an implication of this assumption. Consider a convex set $C$ with a non-empty interior. Then, its centroid $\mathbb{E}[\boldsymbol{M} | \boldsymbol{M} \in C]$ must be in the interior of set $C$.\footnote{This follows from a separating hyperplane argument.} We will use this implication later in the paper.

We formally define a quantization policy in the following. Note that due to results in \citep{CrawfordSobel} and \citep{SaritasTAC2017}, a Nash equilibrium in the scalar source case must involve quantization policies at the encoder with convex bins.

\begin{defn}\label{def:quantizer}
{\normalfont
A quantization policy with $K$ bins, $q$, is a (Borel) measurable mapping from $\mathbb{M}^n=\mathbb{R}^n$ to the set $\{1,\dots,K\}$ characterized by a measurable partition $\{\mathcal{B}^1,\dots,\mathcal{B}^K\}$ such that $\mathcal{B}^i=\{\boldsymbol{m}\,|\,q(\boldsymbol{m})=i\}$ for $i=1,\dots,K$ and that bin probabilities are strictly positive. The $\mathcal{B}^i$ are called the bins of $q$.
}
\end{defn}

The bins defined in Definition~\ref{def:quantizer} lead to a Nash equilibrium under certain conditions described later in the manuscript. If these bins form a Nash equilibrium, they are referred to as (Nash) equilibrium partitions. In contrast to the scalar source case, there may exist a Nash equilibrium with a linear encoder in the multi-dimensional source case, which is investigated later in the paper. Accordingly, we make the following definition.

\begin{defn}\label{def:linearNashEquilibria}
{\normalfont
For the $n$-dimensional cheap talk problem, if an encoding policy $\boldsymbol{z}=\gamma^e(\boldsymbol{m})=A\boldsymbol{m}$ where $A\in \mathbb{R}^{m\times n}$ with $m\leq n$ and a decoding policy $\boldsymbol{u}=\gamma^d(\boldsymbol{z})$ satisfy \eqref{eq:nashEquilibrium}, we say that these policies lead to a linear Nash equilibrium.
}
\end{defn}

\begin{defn}
{\normalfont
We say that a Nash equilibrium is {\it informative} if the encoder reveals information related to the source, i.e., the source $\boldsymbol{M}$ and the message $\boldsymbol{Z}$ are not independent random variables. A Nash equilibrium is referred to as {\it non-informative} when the encoded message is independent of the source.
}
\end{defn}

We note that there always exists a non-informative Nash equilibrium for the multi-dimensional cheap talk problem, which follows from \citep{CrawfordSobel}. In this equilibrium, the encoder transmits a message which is independent of the source. The decoder takes an action based on the prior probability distribution of the source, i.e., its best response $\boldsymbol{u}=\mathbb{E}[\boldsymbol{M}]$. This is a Nash equilibrium since both the encoder and the decoder cannot improve their expected costs by deviating from these strategies. In contrast, an informative Nash equilibrium may or may not exist depending on the setup.

At a given Nash equilibrium, all possible realized values of $\boldsymbol{u}$ are referred to as {\it decoder actions}. While investigating the geometric properties of Nash equilibria, we frequently use the following definition regarding the set of decoder actions in equilibrium.
\begin{defn}
{\normalfont 
We say that a non-empty set of decoder actions containing more than one element forms a {\it continuum} if it is a closed and connected set (i.e., it cannot be expressed as a union of two or more disjoint and closed sets).
}
\end{defn}

An important implication of our results is related to the information theoretic limits of the cheap talk problem. In classical communication settings involving decision makers with aligned goals, information theoretic limits specify bounds on the rate of communication and system performance measured by a common cost criterion (see, e.g., \citep{CoverThomasBook}). In such settings, a bound on the achievable communication rate arises due to system requirements such as a power constraint at the encoder and having a noisy channel. On the other hand, an interesting question arises in a game theoretic setup: Does an upper bound exist on the achievable rate of communication due to misaligned cost criteria? In certain cases, our analysis gives a conclusive answer to this question for the multi-dimensional cheap talk setup. In particular, we show that there exists a Nash equilibrium with a linear encoder depending on certain explicit conditions, in which case there does not exist an upper bound on the achievable rate of communication. We consider the Nash setup for such an information theoretic problem. We refer the reader to \citep{TreustJET2019} for a Stackelberg (Bayesian persuasion) setup.

\subsection{Literature Review}\label{sec:literature}

The cheap talk and signaling game problems have gained significant attention in recent control and communication theory literature. For instance, \citep{SaritasTAC2017} investigates signaling game setups with quadratic cost criteria under Nash and Stackelberg equilibria concepts where a biased encoder communicates with a decoder. The work in \citep{AkyolProcIEEE2017} considers a Gaussian signaling game problem under the Stackelberg equilibrium concept where the bias term at the encoder is modeled as a random variable. In \citep{SayinAutomatica2019}, a multi-stage Gaussian signaling setup is investigated under the Stackelberg equilibrium concept where the private state of the encoder is a controlled Gauss–Markov process. The work in \citep{TreustJET2019} investigates information theoretic limits for the Bayesian persuasion (Stackelberg) setup where there is a commitment assumption for the encoder. The works in \citep{VoraISIT2020} and \citep{VoraCDC2020} consider problems under the Stackelberg equilibrium concept where the decoder has a commitment assumption and introduce the notion of information extraction capacity. In \citep{NumberOfBinsArxiv}, various properties of Nash equilibria are analyzed for the one-dimensional quadratic cheap talk problem. In \citep{SaritasAutomatica2020}, multi-stage cheap talk and signaling game problems are investigated under Nash and Stackelberg equilibria. In \citep{KazikliWiOpt21}, some of the preliminary results in this paper were announced, and the results presented did not include proofs.

Multi-dimensional cheap talk problems have also been considered in the economics literature \citep{LevyRazinEconometrica2007,BattagliniEconometrica2002,MiuraGEB14,ComparativeCheapTalk07,AmbrusTE08}. For instance, \citep{LevyRazinEconometrica2007} investigates a two-dimensional source setting where an encoder communicates with a decoder. Different from our work, the encoder's preferences over different decoder actions are primarily determined by preferences in a certain dimension. In particular, if the encoder prefers one decoder action over the other in this dimension, then the second dimension does not matter. In this case, \citep{LevyRazinEconometrica2007} shows the existence of an upper bound on the number of decoder actions. In addition, the work in \citep{BattagliniEconometrica2002} considers a multi-dimensional cheap talk problem with two encoders and a decoder. While \citep{BattagliniEconometrica2002} studies conditions on the existence of equilibria with the encoders completely revealing their observations, our focus instead is on the characterization of Nash equilibrium partitions in general; as in the case with a single encoder, we do not have full revelation in general. More specifically, we focus on a scenario with a single encoder that jointly encodes its multi-dimensional observation and employs a single quadratic cost function. More recently, \citep{Semirat} investigates a two-dimensional cheap talk setup between an encoder and a decoder considering a uniform source where the encoder is restricted to transmit a binary message. This work proves the existence of an informative Nash equilibrium for any bias vector under the considered setup.

\subsection{Contributions}

The main aim of this paper is to analyze a quadratic multi-dimensional cheap talk problem, which is a multi-dimensional extension of Crawford and Sobel's formulation \citep{CrawfordSobel}. The main contributions of this paper can be summarized as follows:
\begin{enumerate}[(i)]
\item We show that for general source distributions, decoder actions in any Nash equilibrium must satisfy a necessary geometric condition (Lemma~\ref{lem:geoCon}).
\item We derive the necessary conditions that a Nash equilibrium with a continuum of decoder actions needs to satisfy in the case of two-dimensional observations with general distributions (Lemma~\ref{lem:continuum1} and Lemma~\ref{lem:continuum2}).
\item We completely characterize necessary and sufficient conditions under which linear Nash equilibria exist considering independent and identically distributed (i.i.d.) two-dimensional observations (Theorem~\ref{thm:main}). We also generalize these results to the case when the encoder makes more than two i.i.d. observations (Theorem~\ref{thm:gaussian} and Theorem~\ref{thm:sym}).
\item We take the dimension of the source process to infinity and provide an information theoretic perspective to the cheap talk problem by introducing a rate-distortion theoretic formulation. We obtain achievable rates and distortions for the particular case of i.i.d. Gaussian sources (Theorem~\ref{thm:infoTheoretic}).
\end{enumerate}

\section{Geometric Properties of Nash Equilibria}

\subsection{A Necessary Geometric Condition for Nash Equilibria}

In this subsection, we show that the cost structure employed in the problem imposes certain restrictions on the actions taken by the decoder at a Nash equilibrium. In particular, we derive a geometric condition that any two decoder actions at a Nash equilibrium must satisfy. This derivation also allows us to specify the general structure of a Nash equilibrium with a quantization policy at the encoder. In addition, while this geometric condition is important on its own as it provides a necessary condition for a Nash equilibrium in terms of induced decoder actions, it is also useful while deriving conditions for the existence of linear Nash equilibria. It is noted that the following result holds regardless of the source distribution and applies to both i.i.d. and non-i.i.d. sources.

\begin{lem}\label{lem:geoCon}
Consider the $n$-dimensional cheap talk problem with the source random vector $\boldsymbol{M}=[M_1,\dots,M_n]^T$ where each element of $\boldsymbol{M}$ can have different distributions and can be dependent or independent. Let $\mathcal{B}^\alpha$ and $\mathcal{B}^\beta$ be two bins, and let $\boldsymbol{u}^\alpha = \mathbb{E}[\boldsymbol{M} | \boldsymbol{M} \in \mathcal{B}^\alpha]$ and $\boldsymbol{u}^\beta = \mathbb{E}[\boldsymbol{M} | \boldsymbol{M} \in \mathcal{B}^\beta]$ denote their centroids which are the decoder actions taken when the encoder reveals $\boldsymbol{M} \in \mathcal{B}^\alpha$ and $\boldsymbol{M} \in \mathcal{B}^\beta$, respectively. 
\begin{enumerate}[(i)]
\item These decoder actions must satisfy the following necessary condition at a Nash equilibrium:
\begin{align}
2 \, | (\boldsymbol{u}^\beta-\boldsymbol{u}^\alpha)^T \boldsymbol{b} | 
\leq  \lVert \boldsymbol{u}^\beta-\boldsymbol{u}^\alpha \rVert^2. \label{eq:geoCon}
\end{align}
\item At a Nash equilibrium, the encoder decomposes the complete observation space into two regions via a hyperplane orthogonal to $(\boldsymbol{u}^\alpha-\boldsymbol{u}^\beta)$ and intersecting the line connecting $\boldsymbol{u}^\alpha$ and $\boldsymbol{u}^\beta$, and $\mathcal{B}^\alpha$ and $\mathcal{B}^\beta$ are subsets of these respective regions. In particular, $\mathcal{B}^\alpha$ must be a subset of the set $\{\boldsymbol{m} \,|\,h(\boldsymbol{m},\boldsymbol{u}^\alpha,\boldsymbol{u}^\beta)\geq 0 \}$ whereas $\mathcal{B}^\beta$  must be a subset of the set $\{\boldsymbol{m} \,|\,h(\boldsymbol{m},\boldsymbol{u}^\alpha,\boldsymbol{u}^\beta)\leq 0 \}$ where 
\begin{align}
h(&\boldsymbol{m},\boldsymbol{u}^\alpha,\boldsymbol{u}^\beta) \triangleq\nonumber \\
& \bigg(\boldsymbol{m}- \bigg(\frac{\boldsymbol{u}^\beta+\boldsymbol{u}^\alpha}{2}+\boldsymbol{b}\bigg) \bigg)^T (\boldsymbol{u}^\beta-\boldsymbol{u}^\alpha),\label{eq:halfspace}
\end{align}
and $h(\boldsymbol{m},\boldsymbol{u}^\alpha,\boldsymbol{u}^\beta) = 0$ defines the hyperplane on which the encoder is indifferent between either decoder actions, i.e., these $\boldsymbol{m}$ values may belong to both $\mathcal{B}^\alpha$ and $\mathcal{B}^\beta$.
\item At a Nash equilibrium where the encoder uses quantization policies, the quantization bins are always convex. 
\end{enumerate}
\end{lem}

\begin{figure}
\centering
\includegraphics[width=0.8\linewidth]{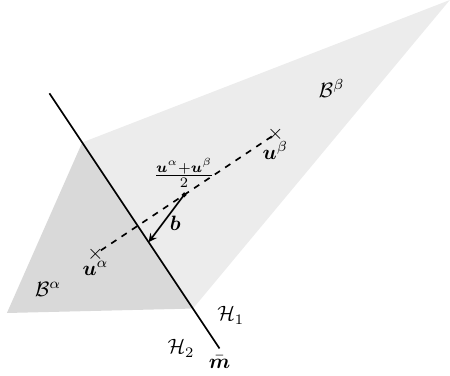}
\caption{Illustration of half spaces induced by decoder actions $\boldsymbol{u}^\alpha$ and $\boldsymbol{u}^\beta$ in Lemma~\ref{lem:geoCon}. The crosses represent the decoder actions, and the arrow represents the bias vector. These decoder actions and the bias vector lead to a line of $\bar{\boldsymbol{m}}$ values for which the encoder is indifferent between reporting these observations as $\boldsymbol{u}^\alpha$ and $\boldsymbol{u}^\beta$. The shaded areas illustrate example bins which satisfy the necessary condition that the half spaces $\mathcal{H}_1$ and $\mathcal{H}_2$ cannot intersect with $\mathcal{B}^\alpha$ and $\mathcal{B}^\beta$, respectively.}
\label{fig:geometry}
\end{figure}

\begin{figure}
\centering
\includegraphics[width=0.3\textwidth]{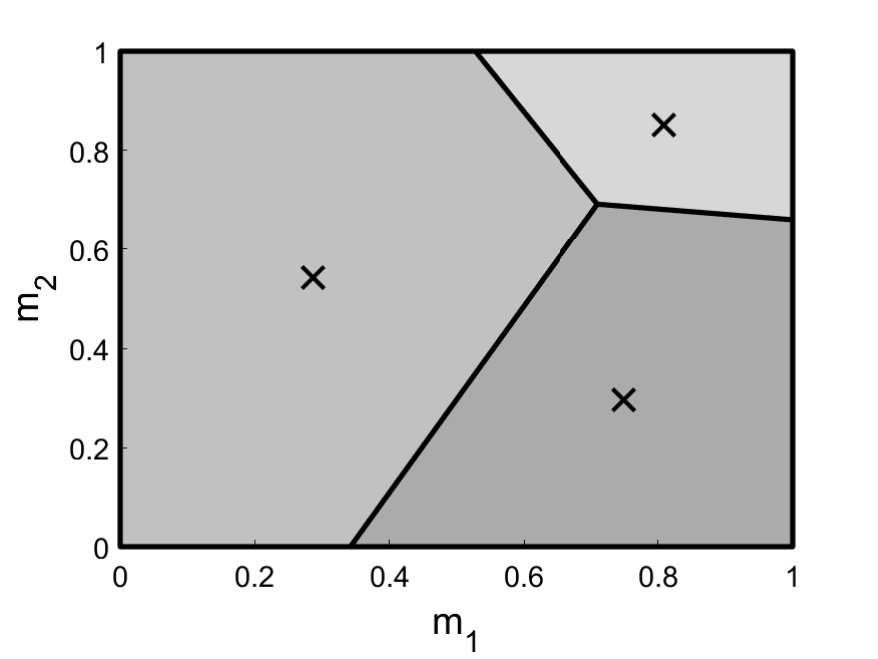}
\caption{Illustration of a Nash equilibrium involving a quantization policy with three bins for the case when $b_1=b_2=0.1$ and the source is two-dimensional i.i.d. with a uniform distribution. The shaded areas show the quantization bins, the lines between the areas are the bin edges, and the crosses represent the decoder actions induced in equilibrium. These quantization bins and bin edges satisfy \eqref{eq:regions} and \eqref{eq:centroidCond}, which leads to a Nash equilibrium.}
\label{fig:quantized}
\end{figure}

See Appendix~\ref{proof:geoCon} for a proof. Fig.~\ref{fig:geometry} illustrates the result in Lemma~\ref{lem:geoCon} for an example setup. In the case of more than two decoder actions, each pair of decoder actions must satisfy the condition in \eqref{eq:geoCon} at a Nash equilibrium. In addition, the bins for each decoder action must be obtained by computing half spaces via \eqref{eq:halfspace} for each pair of decoder actions and then by intersecting these half spaces. In particular, if the decoder actions $\{\boldsymbol{u}^1,\dots,\boldsymbol{u}^K\}$ and the corresponding bins $\{\mathcal{B}^1,\dots,\mathcal{B}^K\}$ form a Nash equilibrium with $K$ bins, then it must be that
\begin{align}
\mathcal{B}^i = \{ \boldsymbol{m} \,|\, h(\boldsymbol{m},\boldsymbol{u}^i,\boldsymbol{u}^j)\geq 0\text{ for all }j\neq i\},\label{eq:regions}
\end{align}
for $i=1,\dots,K$. Note that the conditions in \eqref{eq:regions} are necessary but not sufficient for a Nash equilibrium. Due to the equilibrium conditions at the decoder, for a Nash equilibrium with $K$ bins, the decoder actions $\{\boldsymbol{u}^1,\dots,\boldsymbol{u}^K\}$ and the corresponding bins $\{\mathcal{B}^1,\dots,\mathcal{B}^K\}$ must also satisfy the following centroid conditions:
\begin{align}
\boldsymbol{u}^i = \mathbb{E}[\boldsymbol{M}|\boldsymbol{M}\in \mathcal{B}^i],\label{eq:centroidCond}
\end{align} 
for $i=1,\dots,K$. If the conditions in \eqref{eq:regions} and \eqref{eq:centroidCond} are satisfied, then the corresponding decoder actions and bins form a Nash equilibrium with $K$ bins. Fig.~\ref{fig:quantized} depicts a Nash equilibrium involving a quantization policy with three bins at the encoder.

\begin{rem}
{\normalfont 
In the case of scalar cheap talk, it is required that $|u^\alpha-u^\beta|>2|b|$ holds for any decoder actions $u^\alpha$ and $u^\beta$ at a Nash equilibrium. This directly implies that a Nash equilibrium must involve quantization policies as concluded in \citep[Theorem~3.2]{SaritasTAC2017}. In contrast, such a direct conclusion does not hold for the multi-dimensional cheap talk problem. In fact, if $(\boldsymbol{u}^\beta-\boldsymbol{u}^\alpha)$ is orthogonal to $\boldsymbol{b}$, then the necessary condition in \eqref{eq:geoCon} is always satisfied regardless of the distance between these two decoder actions. This permits the existence of linear Nash equilibria when the source is multi-dimensional, depending on certain conditions investigated later in the paper. Since such an orthogonality property does not hold when the source is one-dimensional, decoder actions cannot get arbitrarily close. Hence, there does not exist a linear Nash equilibrium in this case.
}
\end{rem}

Lemma~\ref{lem:geoCon} presents a geometric condition that any two decoder actions at a Nash equilibrium must satisfy. It is important to emphasize that this condition applies to any joint distribution for multi-dimensional observations. In particular, Lemma~\ref{lem:geoCon} holds even for joint distributions that are not independent and identically distributed.

\subsection{Necessary Conditions for Continuum of Decoder Actions in Equilibrium}\label{sec:equivalent}

In this subsection, we further investigate the geometric condition in Lemma~\ref{lem:geoCon} to derive conditions that a Nash equilibrium with a connected set of decoder actions must satisfy for the particular case of two-dimensional cheap talk. Since a linear encoding policy induces a connected set of decoder actions, our results in this subsection are useful while deriving conditions for the existence of a linear Nash equilibrium. 

\begin{figure}
\centering
\includegraphics[width=\linewidth]{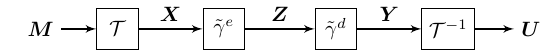}
\caption{Equivalent formulation where $\mathcal{T}$ denotes the linear transformation specified by $X_1=b_1M_2-b_2M_1$ and $X_2=b_1M_1+b_2M_2$, and $\mathcal{T}^{-1}$ denotes its inverse. }
\label{fig:equivalent}
\end{figure}

Lemma~\ref{lem:geoCon} implies that for decoder actions $\boldsymbol{u}^\alpha$ and $\boldsymbol{u}^\beta$ satisfying $ (\boldsymbol{u}^\beta-\boldsymbol{u}^\alpha)^T\boldsymbol{b} = 0$, it is possible to make their distance $\lVert \boldsymbol{u}^\alpha-\boldsymbol{u}^\beta\rVert$ arbitrarily small. On the other hand, for decoder actions $\boldsymbol{u}^\alpha$ and $\boldsymbol{u}^\beta$ with $ (\boldsymbol{u}^\beta-\boldsymbol{u}^\alpha)^T\boldsymbol{b} \neq 0$, since the distance $\lVert \boldsymbol{u}^\alpha-\boldsymbol{u}^\beta\rVert$ is lower bounded by a positive value, these decoder actions $\boldsymbol{u}^\alpha$ and $\boldsymbol{u}^\beta$ cannot get arbitrarily close. This motivates an equivalent formulation by introducing the following transformation of variables. In particular, we define
\begin{align}
&\boldsymbol{X} = \mathcal{T} \boldsymbol{M},\label{eq:MtoX}\\
&\boldsymbol{U} = \mathcal{T}^{-1} \boldsymbol{Y},\label{eq:YtoU}
\end{align}
where 
\begin{align}
\mathcal{T} 
=
\begin{bmatrix}
-b_2 & b_1\\
b_1 & b_2
\end{bmatrix},\quad 
\mathcal{T}^{-1}
=\frac{1}{b_1^2+b_2^2} 
\begin{bmatrix}
-b_2 & b_1\\
b_1 & b_2
\end{bmatrix},
\end{align}
and $\boldsymbol{X}\triangleq [X_1,X_2]^T$ and $\boldsymbol{Y}\triangleq [Y_1,Y_2]^T$ respectively denote the observation at the encoder and the decoder action in the transformed coordinate system. The proposed equivalent formulation is depicted in Fig.~\ref{fig:equivalent} where the linear transformation $\mathcal{T}$ and its inverse $\mathcal{T}^{-1}$ are fixed, and the encoder and decoder design $\tilde{\gamma}^e(\cdot)$ and $\tilde{\gamma}^d(\cdot)$, respectively. In the following lemma, we show that the proposed transformation of variables leads to an equivalent formulation. See Appendix~\ref{proof:equivalent} for a proof. 
\begin{lem}\label{lem:equivalent}
Suppose that the encoder uses a fixed transformation from the source $\boldsymbol{M}$ to an auxiliary variable $\boldsymbol{X}$ specified by \eqref{eq:MtoX} and designs the map $\tilde{\gamma}^e(\cdot)$ from $\boldsymbol{X}$ to the encoded message $\boldsymbol{Z}$. Suppose that the decoder designs the map $\tilde{\gamma}^d(\cdot)$ from its observation $\boldsymbol{Z}$ to an auxiliary variable $\boldsymbol{Y}$ and employs a fixed transformation from $\boldsymbol{Y}$ to the decoder action $\boldsymbol{U}$ specified by \eqref{eq:YtoU}. Then, designing $\tilde{\gamma}^e(\cdot)$ at the encoder and $\tilde{\gamma}^d(\cdot)$ at the decoder is equivalent to the original problem where the encoder designs the map $\gamma^e(\cdot)$ from $\boldsymbol{M}$ to $\boldsymbol{Z}$ under the cost criterion \eqref{eq:encCost}, and the decoder designs the map $\gamma^d(\cdot)$ from $\boldsymbol{Z}$ to $\boldsymbol{U}$ under the cost criterion \eqref{eq:decCost}. In particular, an equilibrium under the proposed formulation is also an equilibrium under the problem given in \eqref{eq:nashEquilibrium} and vice versa. In this equivalent formulation, the aim of the encoder and decoder is to minimize $\tilde{J}^e(\tilde{\gamma}^e,\tilde{\gamma}^d)\triangleq\mathbb{E}[c^e_t(\boldsymbol{X},\boldsymbol{Y})]$ and $\tilde{J}^d(\tilde{\gamma}^e,\tilde{\gamma}^d)\triangleq\mathbb{E}[c^d_t(\boldsymbol{X},\boldsymbol{Y})]$, respectively, where  
\begin{align}
&c^e_t(\boldsymbol{x},\boldsymbol{y}) \triangleq (x_1-y_1 )^2 + (x_2-y_2-\tilde{b})^2 = c^e(\boldsymbol{m},\boldsymbol{u}) \tilde{b},\\
&c^d_t(\boldsymbol{x},\boldsymbol{y}) \triangleq  (x_1-y_1 )^2 + (x_2-y_2)^2 = c^d(\boldsymbol{m},\boldsymbol{u}) \tilde{b},\\
&\tilde{b} \triangleq b_1^2 + b_2^2 \label{eq:defbTilde}.
\end{align}
\end{lem}

\begin{lem}\label{lem:mmse}
For a fixed encoding policy $\tilde{\gamma}^e(\boldsymbol{x})$, the
optimal $\tilde{\gamma}^d(\cdot)$ that minimizes $\tilde{J}^d(\tilde{\gamma}^e,\tilde{\gamma}^d)$ is given by $\mathbb{E}[\boldsymbol{X} |\boldsymbol{Z} = \boldsymbol{z}]$.
\end{lem}

See Appendix~\ref{proof:mmse} for a proof. Equipped with this equivalent formulation, we are now ready to present our results on necessary conditions for any Nash equilibrium with a continuum of decoder actions. 

\begin{lem}\label{lem:continuum1}
Consider the two-dimensional cheap talk problem. Suppose that at a given Nash equilibrium, a set of decoder actions $\mathcal{C}$ forms a continuum. Then, for any $\boldsymbol{y}^\alpha \in \mathcal{C}$ and $\boldsymbol{y}^\beta \in \mathcal{C}$, it must be that $y_2^\alpha=y_2^\beta$.
\end{lem}

See Appendix~\ref{proof:lemmaContinuum1} for a proof. Lemma~\ref{lem:continuum1} implies that a continuum of actions is allowed only in a specific direction that depends on the bias terms in the original coordinate system.

\begin{lem}\label{lem:continuum2}
Consider the two-dimensional cheap talk problem. Suppose that at a given Nash equilibrium, a set of decoder actions with the same second coordinate forms a continuum, i.e., $y_2=\kappa$ where $\kappa$ is in the support of $X_2$. Then, it must be that there exist decoder actions for all values of $y_1\in[x_1^L(\kappa),x_1^U(\kappa)]$ and $y_2=\kappa$ where $x_1^L(\kappa)$ and $x_1^U(\kappa)$ denote lower and upper boundaries of the support of $X_1$ when $X_2=\kappa$, i.e., these decoder actions must be connected. 
\end{lem}

See Appendix~\ref{proof:lemmaContinuum2} for a proof. Lemma~\ref{lem:continuum1} states that a continuum of decoder actions must have a constant $y_2$ coordinate, and Lemma~\ref{lem:continuum2} states that this continuum of decoder actions must be supported \emph{for all} values of $y_1$ in the support of $X_1$ given that $X_2=y_2$. This means that a continuum of decoder actions cannot have a discontinuity. This type of continuum of actions can be attained by revealing the value of $X_1$ completely. In certain scenarios depending on the distribution and the bias vector, revealing $X_1$ can be a Nash equilibrium, as investigated in the next section.

\section{Linear Nash Equilibria}

In this section, we present our main results on the existence of linear Nash equilibria. Towards that goal, we employ Lemma~\ref{lem:continuum1} and Lemma~\ref{lem:continuum2} together with an interesting result from the literature known as Kagan-Linnik-Rao Theorem \citep[Theorem~5.3.1]{KaganBook}. We first consider the two-dimensional case in the following theorem. See Appendix~\ref{proof:main} for a proof. 

\begin{thm}\label{thm:main}
Consider the multi-dimensional cheap talk problem with sources $M_1$ and $M_2$, which are i.i.d. with the corresponding bias terms $b_1$ and $b_2$. 
\begin{enumerate}[(i)]
\item For $b_1=0$ or $b_2=0$, there always exists an informative Nash equilibrium with a linear encoder where the encoder completely reveals the source corresponding to a zero bias.
\item For $b_1\neq 0$, $b_2\neq 0$ and $|b_1|\neq |b_2|$, there exists an informative Nash equilibrium with a linear encoder if and only if the source distribution is Gaussian.
\item For $b_1= b_2\neq 0$, there exists an informative Nash equilibrium with a linear encoder if and only if the source distribution is symmetric around its mean, i.e., denoting the density of $M_1$ by $f(\cdot)$, we have that $f(\mu+x)=f(\mu-x)$ for almost all $x$ where $\mathbb{E}[M_1]=\mu$.
\item For $b_1= -b_2\neq 0$, there always exists an informative Nash equilibrium with a linear encoder regardless of the source distribution. 
\end{enumerate}
\end{thm}

\begin{figure}
\centering
\includegraphics[width=0.3\textwidth]{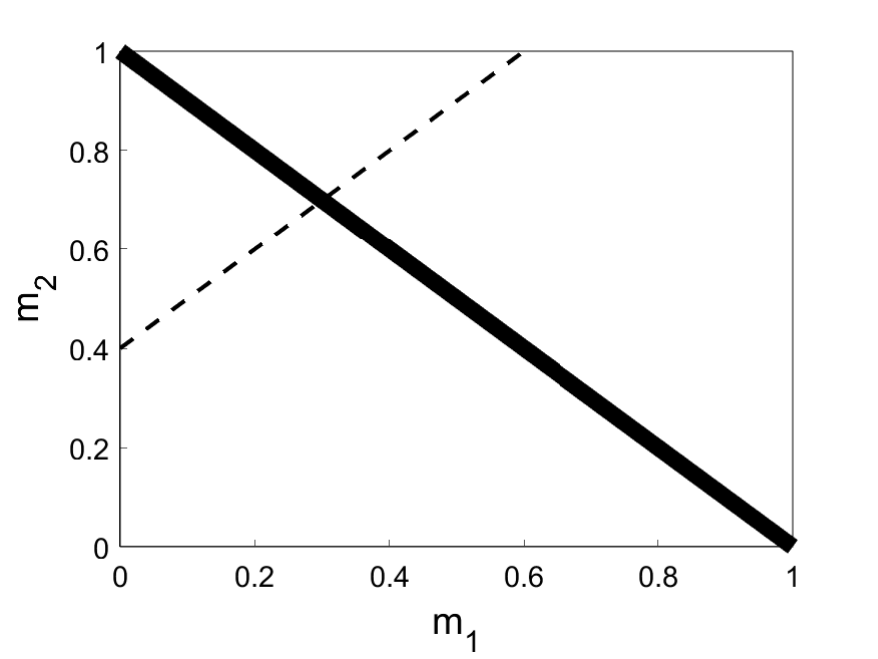}
\caption{Illustration of a Nash equilibrium with a linear encoder $\gamma^e(\boldsymbol{m}) = m_2-m_1$ for the case when $b_1=b_2=0.1$ and the source is two-dimensional i.i.d. with a uniform distribution where the solid line illustrates the continuum of decoder actions induced in equilibrium. The following interpretation can be made in relation to quantization policies (see also Fig.~\ref{fig:quantized}). When the encoder makes an observation exactly on the dashed line $m_2-m_1=0.4$, the encoder only reveals that its observation is on this dashed line. The decoder takes the action $m_1 = 0.3$ and $m_2=0.7$ as its optimal response.}
\label{fig:linear}
\end{figure}

In Figure~\ref{fig:linear}, we depict a linear Nash equilibrium for the case with a two-dimensional uniform source. This scenario corresponds to the third case in Theorem~\ref{thm:main}, where the source distribution is symmetric, and the biases are the same in each dimension.

\begin{rem}
{\normalfont
Lemma~\ref{lem:continuum1} and Lemma~\ref{lem:continuum2} require that at a Nash equilibrium, a continuum of decoder actions can only exist in the direction orthogonal to the bias vector $\boldsymbol{b}$ without any discontinuity considering the original coordinate system. If the source distribution is such that $\mathbb{E}[X_2|X_1=x_1]=\mathbb{E}[b_1M_1+b_2M_2|b_1M_2-b_2M_1=x_1]=0$ holds for all $x_1$, then an encoding policy $z=\gamma^e(\boldsymbol{m})=b_1m_2-b_2m_1$ leads to a continuum of decoder actions that satisfies the necessary conditions in Lemma~\ref{lem:continuum1} and Lemma~\ref{lem:continuum2}. In addition, such an encoding policy leads to a Nash equilibrium, as the proof of Theorem~\ref{thm:main} reveals.    
}
\end{rem}

\begin{rem}
{\normalfont 
Theorem~\ref{thm:main} shows that depending on certain conditions, there exists an informative Nash equilibrium with a linear encoder even for large values of $|b_1|$ and $|b_2|$. On the other hand, in the case of one-dimensional cheap talk, there may exist an upper bound on the number of bins in equilibrium, e.g., for sources with a bounded support \citep{CrawfordSobel} or for log-concave sources with a semi-unbounded support depending on certain conditions \citep{NumberOfBinsArxiv}. In addition, if the bias term is large, this upper bound may even be equal to one, which means that there does not exist an informative Nash equilibrium. Hence, even though the only Nash equilibrium in the case of a one-dimensional scenario may be non-informative, in the case of a two-dimensional scenario with the same bias as in the one-dimensional scenario in both dimensions, it is possible to obtain an informative Nash equilibrium when the source distribution is i.i.d. symmetric.
}
\end{rem}

\begin{rem}
{\normalfont 
In the case of a Gaussian source, the problem decouples into two one-dimensional cheap talk problems. In particular, $X_1=b_1M_2-b_2M_1$ and $X_2=b_1M_1+b_2M_2$ become independent random variables when $M_1$ and $M_2$ are i.i.d. Gaussian. In fact, due to Darmois-Skitovich Theorem \citep[Theorem~3.1.1]{KaganBook}, $X_1$ and $X_2$ are independent only when $M_1$ and $M_2$ are Gaussian. As a result, the problem reduces to obtaining Nash equilibria for decoupled two one-dimensional cheap talk problems where an encoder wishes to convey $X_1$ with a zero bias and another encoder wishes to convey $X_2$ with a bias of $\tilde{b}$. From \citep[Theorem~4]{NumberOfBinsArxiv}, we know that in the case of one-dimensional cheap talk with a Gaussian source, for any $N\geq 1$, there exists a (unique) Nash equilibrium with $N$ bins. Thus, for a two-dimensional cheap talk problem with a Gaussian source, there exists a Nash equilibrium where the encoder reveals $X_1$ completely and applies a quantization policy to $X_2$ with an arbitrary number of bins. 
}
\end{rem}

We can also consider $n$-dimensional i.i.d. Gaussian sources. In this case, one can apply an orthogonal transformation of variables in a similar manner to the two-dimensional case where random variables in each dimension are independent. Under this transformation of variables, there remains a bias term only for a single random variable. Due to \citep[Theorem~4]{NumberOfBinsArxiv} and the independence of the random variables in the transformed coordinate system, it follows that there exists a Nash equilibrium where the encoder applies a quantization policy to this remaining random variable with any number of bins.

\begin{thm}\label{thm:gaussian}
Consider the $n$-dimensional cheap talk problem with an i.i.d. Gaussian source. Then, there exists a Nash equilibrium with a linear encoding policy where the encoder reveals all or a subset of $(n-1)$ dimensions completely (and applies a signaling game policy for the remaining dimension with any number of bins).
\end{thm}
 
See Appendix~\ref{proof:gaussian} for a proof. Theorem~\ref{thm:main} reveals that for the case with an i.i.d. Gaussian source, there always exists a linear Nash equilibrium regardless of the value of the bias vector. One can also consider non-i.i.d. Gaussian sources. In this case, we show that there may exist a linear Nash equilibrium depending on the bias vector and the covariance matrix in the following theorem, whose proof is presented in Appendix~\ref{proof:gaussiancorrelated}.

\begin{thm}\label{thm:gaussiancorrelated}
Consider the multi-dimensional cheap talk problem with Gaussian sources $M_1$ and $M_2$, and the corresponding bias terms $b_1$ and $b_2$. Let $\sigma_1^2$ and $\sigma_2^2$ denote the variances of $M_1$ and $M_2$, respectively, and let $\rho$ denote their covariance. Then, there exists a Nash equilibrium with a linear encoding policy if $b_1b_2 (\sigma_2^2-\sigma_1^2) + (b_1^2-b_2^2) \rho =0$ holds.
\end{thm}

Theorem~\ref{thm:main} reveals that for the case with an i.i.d. two-dimensional symmetric source, there always exists an informative linear Nash equilibrium. When $n>2$, it is possible to apply the linear policy in Theorem~\ref{thm:main} for pairs of random variables to obtain a linear Nash equilibrium as we can obtain decoupled two-dimensional cheap talk problems. For instance, if $n=2k$ for some $k>2$, then revealing all or a subset of the random variables $M_2-M_1$, $M_4-M_3$, $\dots$, $M_{2k}-M_{2k-1}$ yields a Nash equilibrium. In this case, the encoder reveals at most $n/2$ dimensions. In the following theorem, we show that a joint encoding policy can be applied to obtain a Nash equilibrium where the encoder reveals $(n-1)$ dimensions. The proof appears in Appendix~\ref{proof:sym}.

\begin{thm}\label{thm:sym}
Consider the $n$-dimensional cheap talk problem involving an i.i.d. source with a symmetric distribution. Then, there exists a Nash equilibrium with a linear encoding policy where the encoder reveals all or a subset of $(n-1)$ dimensions completely in a transformed coordinate system.
\end{thm}

\section{Large Dimensions and a Rate-Distortion Theoretic Formulation of Cheap Talk} 

We have analyzed the multi-dimensional cheap talk problem where the bias vector at the encoder can be arbitrary. The special case when the components of the bias vector are the same leads to an important problem from an information theoretic perspective. In this case, the problem is to convey an i.i.d. source with a certain bias, and the bias is the same for each source component. In other words, the encoder observes independent copies from a random source and wishes to introduce the same bias for each independent copy. In such a problem, one may wish to obtain information theoretic limits of the communication. In a sense, this problem is a game theoretic counterpart of rate-distortion theory that is studied in a classical communication theoretic setup. Our findings reveal that if the distribution is Gaussian, then there always exists a linear Nash equilibrium where the encoder completely reveals $(n-1)$ dimensions in a transformed coordinate system. For the remaining dimension, the encoder has to employ a signaling game policy with an arbitrary number of bins, including the case with one bin. This result holds because the problem can be transformed into decoupled problems consisting of a team theoretic problem for conveying an $(n-1)$-dimensional i.i.d. source without any bias and a one-dimensional cheap talk problem with a certain bias in the remaining dimension. If we increase the number of observed sources at the encoder, the effect of employed policy for this remaining dimension becomes negligible. This implies that the problem of finding achievable rate and distortion pairs is asymptotically equivalent to obtaining achievable rate and distortion pairs for a team theoretic setup in a transformed coordinate system.

The problem of interest is in fact can be more generally expressed in a rate-distortion theoretic formulation. The aim is to find the achievable rate and distortion region. In particular, we have the following problem:
\begin{prob}\label{problemRateDistortion}
{\normalfont 
Consider the $n$-dimensional cheap talk problem with i.i.d. sources and $b=b_1=\dots=b_n$. We say that a tuple of rate and distortion pairs $(R,D_e,D_d)$ is {\it achievable} at a Nash equilibrium if there exists a sequence of encoders and decoders that leads to a Nash equilibrium with the following properties:
\begin{enumerate}[(i)]
\item The encoder is given by $\gamma^e_n : \mathbb{M}^n\to \{1,\dots,2^{nR}\}$. 
\item The decoder is given by $\gamma^d_n : \{1,\dots,2^{nR}\} \to \mathbb{M}^n$ such that 
\begin{align}
&\lim_{n\to\infty}\frac{\mathbb{E}\left[\sum_{i=1}^n(M_i-U_i-b)^2\right]}{n} \leq D_e, \label{eq:DeConstraint}\\
&\lim_{n\to\infty} \frac{\mathbb{E}\left[\sum_{i=1}^n(M_i-U_i)^2\right]}{n} \leq D_d.\label{eq:DdConstraint}
\end{align}
\end{enumerate}
Then, the problem is to determine if a given tuple $(R,D_e,D_d)$ is {\it achievable} at a Nash equilibrium. 
}
\end{prob}

If the bias term is zero in this problem, then we obtain a team theoretic problem since the corresponding distortion values are identical at the encoder and decoder. We denote the corresponding rate and distortion values by $R_T$ and $D_T$, respectively, where the subscript refers to the fact that the setup is team theoretic.

While we leave the study of Problem~\ref{problemRateDistortion} for general sources for future work, the Gaussian case is completely solvable. Our result in the previous section shows that if the source distribution is Gaussian, one can apply a suitable transformation of variables to obtain an equivalent problem for which the encoder has a bias only for a single random variable. We use this idea to relate achievable rate and distortion values of the original problem to that of a team theoretic problem. Before presenting this result, we note that at a Nash equilibrium we have $\mathbb{E}[\sum_{i=1}^n(M_i-U_i-b)^2]=\mathbb{E}[\sum_{i=1}^n(M_i-U_i)^2]+b^2n$. Thus, we have the same rate region for any $D_e$ value satisfying $D_e\geq D_d+b^2$. The following theorem characterizes achievable rates and distortion values for Problem~\ref{problemRateDistortion} with Gaussian sources. See Appendix~\ref{proof:infoTheoretic} for a proof.

\begin{thm}\label{thm:infoTheoretic}
Consider the multi-dimensional cheap talk problem with i.i.d. Gaussian sources where the bias term $b$ is the same at each dimension. Suppose that a rate and a distortion pair $(R_T,D_T)$ is achievable for the team theoretic problem with a zero bias. Then, for the game theoretic problem with a non-zero bias, the following rate and distortion values are achievable:
\begin{align}
R = R_T,\quad 
D_e \geq D_T+b^2,\quad 
D_d \geq D_T.
\end{align}
\end{thm}

\begin{rem}
{\normalfont 
The proof of Theorem~\ref{thm:gaussian} reveals that by applying a joint encoding policy that uses multi-dimensional observations, it is possible to achieve team theoretic rates and distortions (except that there is still an additional $b^2$ term for the encoder's distortion). If we do not allow for joint encoding, which is equivalent to considering the scalar cheap talk setup, then the same rate in the team setup and in the game setup leads to different distortion values at the decoder, with the latter being larger.
}
\end{rem}

In rate-distortion theory, an important concept is the rate-distortion function. In a classical communication theoretic setup, this is defined as the infimum of rates $R$ such that $(R,D)$ is achievable. A similar definition of rate-distortion function in a game theoretic setup yields
\begin{align}
R(D_e,D_d) \triangleq \inf\{ R \,|\, (R,D_e,D_d) \text{ is achievable} \}.
\end{align}
By using the result of Theorem~\ref{thm:infoTheoretic}, we can upper bound the rate-distortion function for the Gaussian case. Towards that goal, we use the following result for the team theoretic setup where the distortion $D$ is identical at the encoder and decoder as there is no bias at the encoder.

\begin{lem}\label{thm:RDTeam} \citep[Theorem~10.3.2]{CoverThomasBook}
{
Suppose that $b=0$ in Problem~\ref{problemRateDistortion}. Consider i.i.d. Gaussian sources with a variance of $\sigma^2$. The rate distortion function for such a setup is given by $ R(D) = 
\frac{1}{2} \log_2\frac{\sigma^2}{D}$ if $0\leq D \leq \sigma^2$ and $R(D)=0$ if $D> \sigma^2$.
}
\end{lem}

Next, the following theorem, whose proof appears in Appendix~\ref{proof:RDGame}, presents our result for the game theoretic setup with a biased encoder.

\begin{thm}\label{thm:RDGame}
Consider the multi-dimensional cheap talk problem with i.i.d. Gaussian sources where the bias term $b$ is the same at each dimension. The rate-distortion function for such a setup satisfies $R(D_e,D_d) \leq 
\frac{1}{2} \log_2\frac{\sigma^2}{\min\{D_d,D_e-b^2\}}$ if $0\leq \min\{D_d,D_e-b^2\}\leq \sigma^2$, and $R(D_e,D_d)=0$ if $\min\{D_d,D_e-b^2\}> \sigma^2$.
\end{thm}

\begin{rem}
{\normalfont 
A related result can be found in \citep[Theorem~8]{NumberOfBinsArxiv}. It is shown that having more bins in the quantized encoding policy leads to reduced distortion values if the scalar source has a log-concave distribution. In other words, a large rate leads to smaller expected costs for both players under a log-concave source assumption, which holds for the Gaussian case.
}
\end{rem}

\begin{rem} 
{\normalfont 
For multi-dimensional i.i.d. Gaussian sources, there exists a Nash equilibrium where the encoder reveals $(n-1)$ dimensions and applies a signaling game policy for the remaining dimension $X_n$ with an arbitrary number of bins. Since the Gaussian distribution is log-concave, from \citep[Theorem~8]{NumberOfBinsArxiv}, it follows that the expected costs of both players reduce when the number of bins for the quantization policy applied to $X_n$ is increased. In addition, it is also possible to have a Nash equilibrium with infinitely many bins applied to $X_n$ due to \citep[Theorem~13]{NumberOfBinsArxiv}. Hence, a Nash equilibrium where $X_1,\dots,X_{n-1}$ are revealed and a quantization policy with infinitely many bins applied to $X_n$ corresponds to a payoff dominant Nash equilibrium \citep{HarsanyiSeltenBook1988}.
}
\end{rem}

\section{Conclusion}\label{sec:conclusion} 
We have analyzed a quadratic multi-dimensional cheap talk problem. First, we have derived the necessary general conditions for a Nash equilibrium considering any joint source distribution. In particular, we have shown that decoder actions at a Nash equilibrium need to satisfy a geometric condition that essentially prevents any two decoder actions from being arbitrarily close to each other depending on their difference as vectors and the bias vector. Then, we have investigated continuum of decoder actions considering two-dimensional sources and provided a condition that a continuum of decoder actions must satisfy in any Nash equilibrium. Then, we have derived necessary and sufficient conditions under which a linear Nash equilibrium exists considering i.i.d. sources. These conditions require a Gaussian or a symmetric source density. Moreover, we have formulated a rate-distortion theoretic problem for the cheap talk setup and have solved the Gaussian case.

\begin{appendices}

\section{Proof of Lemma~\ref{lem:geoCon}}\label{proof:geoCon} 

\begin{enumerate}[(i)]
\item Let there be two bins, $\mathcal{B}^\alpha$ and $\mathcal{B}^\beta$. Denote their centroids by $\boldsymbol{u}^\alpha = \mathbb{E}[\boldsymbol{M} | \boldsymbol{M} \in \mathcal{B}^\alpha]$ and $\boldsymbol{u}^\beta = \mathbb{E}[\boldsymbol{M} | \boldsymbol{M} \in \mathcal{B}^\beta]$. The encoder is indifferent between the decoder actions $\boldsymbol{u}^\alpha$ and $\boldsymbol{u}^\beta$ for source observation values $\bar{\boldsymbol{m}}$ which satisfy the following:
\begin{align}
& c^e(\bar{\boldsymbol{m}},\boldsymbol{u}^\alpha) = c^e(\bar{\boldsymbol{m}},\boldsymbol{u}^\beta)\nonumber\\
\Leftrightarrow\,
&\lVert \bar{\boldsymbol{m}}-\boldsymbol{u}^\alpha -\boldsymbol{b} \rVert^2 
= \lVert \bar{\boldsymbol{m}}-\boldsymbol{u}^\beta -\boldsymbol{b} \rVert^2\nonumber\\
\Leftrightarrow\,
& (2\bar{\boldsymbol{m}}- (\boldsymbol{u}^\beta+\boldsymbol{u}^\alpha+2\boldsymbol{b}) )^T(\boldsymbol{u}^\beta-\boldsymbol{u}^\alpha) = 0.\label{eq:indifferent}
\end{align}
In other words, if an observation satisfies \eqref{eq:indifferent}, the encoder's costs are the same under the decoder actions $\boldsymbol{u}^\alpha$ and $\boldsymbol{u}^\beta$. From \eqref{eq:indifferent}, it is seen that these $\bar{\boldsymbol{m}}$ values define a hyperplane orthogonal to $(\boldsymbol{u}^\beta-\boldsymbol{u}^\alpha)$. Given any source observation $\boldsymbol{m} = \bar{\boldsymbol{m}} + \Delta (\boldsymbol{u}^\beta-\boldsymbol{u}^\alpha)$ with $\Delta>0$ where $\bar{\boldsymbol{m}}$ satisfies \eqref{eq:indifferent}, the encoder prefers the decoder action $\boldsymbol{u}^\beta$ over the decoder action $\boldsymbol{u}^\alpha$ since the following holds:
\begin{align}
& c^e(\bar{\boldsymbol{m}}+\Delta(\boldsymbol{u}^\beta-\boldsymbol{u}^\alpha),\boldsymbol{u}^\beta)  
-c^e(\bar{\boldsymbol{m}}+\Delta(\boldsymbol{u}^\beta-\boldsymbol{u}^\alpha),\boldsymbol{u}^\alpha) \nonumber \\
&=
- 2\Delta \lVert \boldsymbol{u}^\beta-\boldsymbol{u}^\alpha\rVert^2 <0.
\end{align}
This implies that $\mathcal{B}^\alpha$ and $\mathcal{H}_1$ are disjoint sets where
\begin{align}
\mathcal{H}_1\triangleq 
&\{\boldsymbol{m}\,|\, \boldsymbol{m}= \bar{\boldsymbol{m}} + \Delta (\boldsymbol{u}^\beta-\boldsymbol{u}^\alpha) \nonumber\\
&\;\;\,\text{where }\bar{\boldsymbol{m}}\text{ satisfies \eqref{eq:indifferent} and }\Delta>0\}.\label{eq:H1}
\end{align}
Similarly, given any source observation $\boldsymbol{m} = \bar{\boldsymbol{m}} + \Delta (\boldsymbol{u}^\beta-\boldsymbol{u}^\alpha)$ with $\Delta<0$, the encoder prefers the decoder action $\boldsymbol{u}^\alpha$ over the decoder action $\boldsymbol{u}^\beta$. It follows that $\mathcal{B}^\beta$ and $\mathcal{H}_2$ are disjoint sets where
\begin{align}
\mathcal{H}_2\triangleq 
&\{\boldsymbol{m}\,|\, \boldsymbol{m}= \bar{\boldsymbol{m}} + \Delta (\boldsymbol{u}^\beta-\boldsymbol{u}^\alpha) \nonumber\\
&\;\;\,\text{where }\bar{\boldsymbol{m}}\text{ satisfies \eqref{eq:indifferent} and }\Delta<0\}.\label{eq:H2}
\end{align}
 
Furthermore, the plane specified by \eqref{eq:indifferent} intersects the affine set $\lambda \boldsymbol{u}^\beta+(1-\lambda)\boldsymbol{u}^\alpha$ with $\lambda\in\mathbb{R}$ at a single point due to the fact that $(\boldsymbol{u}^\beta-\boldsymbol{u}^\alpha)$ and the hyperplane specified by \eqref{eq:indifferent} are orthogonal. In order to find the value of $\lambda$ that gives this intersection point, we solve \eqref{eq:indifferent} and $\bar{\boldsymbol{m}}=\bar{\lambda} \boldsymbol{u}^\beta+(1-\bar{\lambda})\boldsymbol{u}^\alpha$ together and obtain an expression for $\bar{\lambda}$ in the following: 
\begin{align}
&\big(2\bar{\boldsymbol{m}}- \boldsymbol{u}^\beta-\boldsymbol{u}^\alpha-2\boldsymbol{b} \big)^T
\big(\boldsymbol{u}^\beta-\boldsymbol{u}^\alpha\big) = 0\nonumber \\
\Leftrightarrow
&(2\bar{\lambda}-1) = \frac{2(\boldsymbol{u}^\beta-\boldsymbol{u}^\alpha)^T\boldsymbol{b}}{\lVert\boldsymbol{u}^\beta-\boldsymbol{u}^\alpha\rVert^2}.\label{eq:lambdaBar}
\end{align}
We know that the centroid conditions require $\boldsymbol{u}^\alpha = \mathbb{E}[\boldsymbol{M} | \boldsymbol{M} \in \mathcal{B}^\alpha]$ and $\boldsymbol{u}^\beta = \mathbb{E}[\boldsymbol{M} | \boldsymbol{M} \in \mathcal{B}^\beta]$. Since $\mathcal{B}^\alpha$ and $\mathcal{H}_1$ are disjoint sets, and $\mathcal{B}^\beta$ and $\mathcal{H}_2$ are disjoint sets, we need $0\leq \bar{\lambda} \leq 1$, which is equivalent to $|2\bar{\lambda}-1|\leq1$. By combining this inequality with \eqref{eq:lambdaBar}, it follows that \eqref{eq:geoCon} holds.
\item As noted above, $\mathcal{B}^\alpha$ and $\mathcal{B}^\beta$ do not intersect with $\mathcal{H}_1$ and $\mathcal{H}_2$, respectively, where $\mathcal{H}_1$ and $\mathcal{H}_2$ are defined in \eqref{eq:H1} and \eqref{eq:H2}. These regions lead to the decomposition specified by the hyperplane in \eqref{eq:halfspace}, which is orthogonal to $(\boldsymbol{u}^\beta-\boldsymbol{u}^\alpha)$.
\item The decision regions for two decoder actions must be constructed by computing and intersecting half spaces. Since a half space is a convex set and intersection operation preserves convexity, the quantization bins must be convex \citep[p.~36]{BoydBook}.
\end{enumerate}

\section{Proof of Lemma~\ref{lem:equivalent}}\label{proof:equivalent} 
For the cost function of the encoder, we can write
\begin{align}
&c^e(\boldsymbol{m},\boldsymbol{u}) 
= (\boldsymbol{m}-\boldsymbol{u}-\boldsymbol{b})^T (\boldsymbol{m}-\boldsymbol{u}-\boldsymbol{b}) \nonumber \\
&= 
\big(\mathcal{T}^{-1}\mathcal{T}(\boldsymbol{m}-\boldsymbol{u}-\boldsymbol{b})\big)^T
\big(\mathcal{T}^{-1}\mathcal{T}(\boldsymbol{m}-\boldsymbol{u}-\boldsymbol{b})\big) \nonumber \\
&= 
\big(\mathcal{T}(\boldsymbol{m}-\boldsymbol{u}-\boldsymbol{b})\big)^T
(\mathcal{T}^{-1})^T
(\mathcal{T}^{-1})
\big(\mathcal{T}(\boldsymbol{m}-\boldsymbol{u}-\boldsymbol{b})\big) \nonumber \\
&=\tilde{b}^{-1}\big(\boldsymbol{x}-\boldsymbol{y}-[0,\,\tilde{b}]^T\big)^T
\big(\boldsymbol{x}-\boldsymbol{y}-[0,\,\tilde{b}]^T\big) \nonumber \\
&=c^e_t(\boldsymbol{x},\boldsymbol{y})\tilde{b}^{-1}, \label{eq:defCostEncTra}
\end{align}
where $\tilde{b}$ is specified in \eqref{eq:defbTilde}, and the fourth equation uses $(\mathcal{T}^{-1})^T
(\mathcal{T}^{-1}) = \tilde{b}^{-1} I$ with $I$ denoting identity matrix, $\mathcal{T}\boldsymbol{b} = [0,\,\tilde{b}]^T$, $\mathcal{T}\boldsymbol{m} = \boldsymbol{x}$ and $\mathcal{T}\boldsymbol{u} = \boldsymbol{y}$. In a similar manner, the cost function of the decoder can be expressed as
\begin{align}
&c^d(\boldsymbol{m},\boldsymbol{u})  
= (\boldsymbol{m}-\boldsymbol{u})^T (\boldsymbol{m}-\boldsymbol{u}) \nonumber \\
&= \big(\mathcal{T}(\boldsymbol{m}-\boldsymbol{u})\big)^T
(\mathcal{T}^{-1})^T
(\mathcal{T}^{-1})
\big(\mathcal{T}(\boldsymbol{m}-\boldsymbol{u})\big) \nonumber \\
&=\tilde{b}^{-1}\big(\boldsymbol{x}-\boldsymbol{y}\big)^T
\big(\boldsymbol{x}-\boldsymbol{y}\big)  =c^d_t(\boldsymbol{x},\boldsymbol{y})\tilde{b}^{-1}, \label{eq:defCostDecTra}
\end{align}
where the third equation uses $(\mathcal{T}^{-1})^T
(\mathcal{T}^{-1}) = \tilde{b}^{-1} I$, $\mathcal{T}\boldsymbol{m} = \boldsymbol{x}$ and $\mathcal{T}\boldsymbol{u} = \boldsymbol{y}$. Note that the factor of $(1/\tilde{b})$ is canceled in the definitions of $c^e_t(\boldsymbol{x},\boldsymbol{y})$ and $c^d_t(\boldsymbol{x},\boldsymbol{y})$ for notational convenience. Since $\tilde{b}$ is a constant and $\tilde{b}>0$, this cancellation does not change the goals of the players. 

Note that a fixed and invertible transformation of observation and action variables independent of the policies $\tilde{\gamma}^e(\cdot)$ and $\tilde{\gamma}^d(\cdot)$ is considered. Moreover, since the transformation is invertible, there is no loss of information at the encoder and the decoder due to the transformation. If $\gamma^{*,e}(\cdot)$ and $\gamma^{*,d}(\cdot)$ satisfy \eqref{eq:nashEquilibrium}, then $\tilde{\gamma}^{*,e}(\cdot) = \gamma^{*,e}(\mathcal{T}^{-1}(\cdot))$ and $\tilde{\gamma}^{*,d}(\cdot) =\mathcal{T}(\gamma^{*,d}(\cdot))$ satisfy $\tilde{J}^e(\tilde{\gamma}^{*,e}, \tilde{\gamma}^{*,d}) \leq \tilde{J}^e(\tilde{\gamma}^{e}, \tilde{\gamma}^{*,d})$ for all $\tilde{\gamma}^e$ and $\tilde{J}^d(\tilde{\gamma}^{*,e}, \tilde{\gamma}^{*,d}) \leq \tilde{J}^d(\tilde{\gamma}^{*,e}, \tilde{\gamma}^{d})$ for all $\tilde{\gamma}^d$ since the cost functions in the transformed and original formulations are essentially the same as shown in \eqref{eq:defCostEncTra} and \eqref{eq:defCostDecTra}. Similarly, if $\tilde{\gamma}^{*,e}(\cdot) $ and $\tilde{\gamma}^{*,d}(\cdot)$ satisfy $\tilde{J}^e(\tilde{\gamma}^{*,e}, \tilde{\gamma}^{*,d}) \leq \tilde{J}^e(\tilde{\gamma}^{e}, \tilde{\gamma}^{*,d})$ for all $\tilde{\gamma}^e$ and $\tilde{J}^d(\tilde{\gamma}^{*,e}, \tilde{\gamma}^{*,d}) \leq \tilde{J}^d(\tilde{\gamma}^{*,e}, \tilde{\gamma}^{d})$ for all $\tilde{\gamma}^d$, then $\gamma^{*,e}(\cdot) = \tilde{\gamma}^{*,e}(\mathcal{T}(\cdot))$ and $\gamma^{*,d}(\cdot) = \mathcal{T}^{-1}(\tilde{\gamma}^{*,d}(\cdot))$ satisfy \eqref{eq:nashEquilibrium}. These reveal that an equilibrium under the proposed (original) formulation is also an equilibrium under the original (proposed) formulation. This equivalence can be viewed as a special case of the result in \citep[Theorem~3.1]{SinaArxiv} where it is shown that for a dynamic stochastic game setup one can equivalently consider its policy-independent static reduction under the Nash equilibrium concept.

\section{Proof of Lemma~\ref{lem:mmse}}\label{proof:mmse} 

For a given encoding policy, the aim of the decoder is to minimize $\tilde{J}^d(\tilde{\gamma}^e,\tilde{\gamma}^d)=\mathbb{E}[c^d_t(\boldsymbol{X},\boldsymbol{Y})]$ where $c^d_t(\cdot,\cdot)$ is as in \eqref{eq:defCostDecTra}. Since the expression in \eqref{eq:defCostDecTra} involves a sum of squared error terms, the result immediately follows.

\section{Proof of Lemma~\ref{lem:continuum1}}\label{proof:lemmaContinuum1} 

This result is a consequence of Lemma~\ref{lem:geoCon}. By using the cost function of the encoder and the decoder considering the equivalent formulation, it can be shown that the condition in \eqref{eq:geoCon} translates to the condition that for any decoder actions $\boldsymbol{y}^\alpha=[y_1^\alpha,y_2^\alpha]^T$ and $\boldsymbol{y}^\beta=[y_1^\beta,y_2^\beta]^T$, we have that
\begin{align}
0 \leq  (y_1^\alpha-y_1^\beta)^2 + (y_2^\alpha-y_2^\beta)^2-2 \tilde{b}|y_2^\alpha-y_2^\beta| \triangleq g(\boldsymbol{y}^\alpha,\boldsymbol{y}^\beta). 
\label{eq:geoConTransformed}
\end{align}
Let there be a continuum of decoder actions that does not have a constant $y_2$ coordinate. Since a continuum of decoder actions with a constant $y_1$ coordinate is not allowed due to \eqref{eq:geoConTransformed}, it is possible to partition a continuum of actions so that it consists of representations of the form $\boldsymbol{y}=[y_1,h_i(y_1)]^T$ for some continuous functions $h_i(\cdot)$ with $i\in\{1,\dots,k\}$. Here, we represent the second coordinate of the continuum as a function of the first coordinate. In the following, we take such a continuum denoted by $\boldsymbol{y}=[y_1,h(y_1)]^T$ for some continuous function $h(\cdot)$ and prove that it leads to a contradiction when $h(\cdot)$ is not a constant function.

Suppose, by contradiction, that $h(\cdot)$ is not a constant function. It follows that we can find two decoder actions $\boldsymbol{y}^\alpha=[y_1^\alpha,y_2^\alpha]^T$ and $\boldsymbol{y}^\beta=[y_1^\beta,y_2^\beta]^T$ on the continuum with the property that $y_2^\alpha\neq y_2^\beta$ and that $y_1^\alpha< y_1^\beta$. Without loss of generality, take $y_2^\alpha< y_2^\beta$. In the following, we first prove the result by making the additional assumption that $h(\cdot)$ is differentiable. By this assumption, we are able to invoke the standard mean value theorem of calculus to conclude the result. However, the result holds also for a non-differentiable $h(\cdot)$. We first prove the result for a differentiable $h(\cdot)$ since this proof is more intuitive. Then, we prove the result in the general case when $h(\cdot)$ is non-differentiable.

Let $h(\cdot)$ be differentiable. By the mean value theorem of calculus, there exists $y_1^\gamma\in[y_1^\alpha,y_1^\beta]$ such that $h'(y_1^\gamma)=(y_2^\beta-y_2^\alpha)/(y_1^\beta-y_1^\alpha)>0$. In particular, we can find a decoder action $\boldsymbol{y}^\gamma$ on the continuum with $y_1^\gamma\in[y_1^\alpha,y_1^\beta]$ and $y_2^\gamma=h(y_1^\gamma)$ such that the derivative of $h(\cdot)$ at $y_1^\gamma$ is strictly positive. Next, we take another decoder action $\boldsymbol{y}^\eta=[y_1^\eta,h(y_1^\eta)]^T$ on the continuum and vary its first coordinate to reach a contradiction to \eqref{eq:geoConTransformed}. In particular, if we express the condition imposed by \eqref{eq:geoConTransformed} for the decoder action $\boldsymbol{y}^\gamma$ and a decoder action on the continuum denoted by $\boldsymbol{y}^\eta$, we get
\begin{align*}
g(\boldsymbol{y}^\eta,\boldsymbol{y}^\gamma) &=(y_1^\eta-y_1^\gamma)^2 \\
&+ (h(y_1^\eta)-h(y_1^\gamma))^2-2 \tilde{b}| h(y_1^\eta)-h(y_1^\gamma)| \geq 0.
\end{align*}
When $y_1^\eta>y_1^\gamma$, and $(y_1^\eta-y_1^\gamma)$ is sufficiently small, we have $h(y_1^\eta)>h(y_1^\gamma)$ due to a positive derivative at $y_1^\gamma$. For fixed $y_1^\gamma$, if we take the derivative of $g(\boldsymbol{y}^\eta,\boldsymbol{y}^\gamma)$ with respect to $y_1^\eta$, we get
\begin{align}
\frac{d g(\boldsymbol{y}^\eta,\boldsymbol{y}^\gamma)}{d y_1^\eta}&=2(y_1^\eta-y_1^\gamma) \nonumber \\
&+2 (h(y_1^\eta)-h(y_1^\gamma))h'(y_1^\eta) - 2\tilde{b}h'(y_1^\eta).\label{eq:lemmaDerivative}
\end{align}
If we take $y_1^\eta=y_1^\gamma$ in \eqref{eq:lemmaDerivative}, then the first two terms are zero while the third term is negative since $h'(y_1^\gamma)>0$ and $\tilde{b}>0$. Therefore, we have that $\left.\frac{d g(\boldsymbol{y}^\eta,\boldsymbol{y}^\gamma)}{d y_1^\eta}\right\vert_{y_1^\eta=y_1^\gamma}<0$. This is a contradiction to $g(\boldsymbol{y}^\eta,\boldsymbol{y}^\gamma)\geq 0$ since $g(\boldsymbol{y}^\gamma,\boldsymbol{y}^\gamma)=0$ and $\left.\frac{d g(\boldsymbol{y}^\eta,\boldsymbol{y}^\gamma)}{d y_1^\eta}\right\vert_{y_1^\eta=y_1^\gamma}<0$. Therefore, it must be that $h(\cdot)$ is a constant function, which means that a continuum of actions must have a constant $y_2$ coordinate. 

Now, consider the general case when $h(\cdot)$ is not differentiable. By the mean value theorem in \citep[Corollary~1]{HiriartUrruty1980}, there exists $y_1^*\in[y_1^\alpha,y_1^\beta]$ such that either $s^*\in \partial h(y_1^*)$ or $s^*\in {-\partial}{(-h(y_1^*))}$ holds where $s^*\triangleq (y_2^\beta-y_2^\alpha)/(y_1^\beta-y_1^\alpha)>0$, and $\partial h(y_1^*)$ denotes the set of subgradients of $h(\cdot)$ at $y_1^*$. We take two decoder actions and employ the condition in \eqref{eq:geoConTransformed} to reach a contradiction. Let $\boldsymbol{y}^\gamma$ and $\boldsymbol{y}^\eta$ denote these two decoder actions, which are expressed as $y_1^\gamma=y_1$, $y_2^\gamma=h(y_1^\gamma)$, $y_1^\eta=y_1+td'$ and $y_2^\eta=h(y_1^\eta)$. If we express the condition imposed by \eqref{eq:geoConTransformed} for the decoder actions $\boldsymbol{y}^\gamma$ and $\boldsymbol{y}^\eta$, we get
\begin{align}
&g(\boldsymbol{y}^\eta,\boldsymbol{y}^\gamma) \nonumber\\
&=(y_1^\eta-y_1^\gamma)^2 + (h(y_1^\eta)-h(y_1^\gamma))^2-2 \tilde{b}| h(y_1^\eta)-h(y_1^\gamma)| \nonumber \\
&= (td')^2 +(h(y_1+td')-h(y_1))^2 
\nonumber\\
&\hphantom{PHANTOM }
-2 \tilde{b}| h(y_1+td')-h(y_1)| \label{eq:lem3ineq1}
\geq 0.
\end{align}
If $t$ and $d'$ are positive and sufficiently small, the inequality in \eqref{eq:lem3ineq1} can be expressed as
\begin{align}
|h(y_1+td')-h(y_1)| + \sqrt{\tilde{b}^2-(td')^2} - \tilde{b}\leq 0.\label{eq:lem3ineq2}
\end{align} 
Since $t$ is positive, it follows from \eqref{eq:lem3ineq2} that
\begin{align}
\frac{|h(y_1+td')-h(y_1)|}{t} + \frac{\sqrt{\tilde{b}^2-(td')^2} - \sqrt{\tilde{b}^2}}{t}\leq 0.\label{eq:lem3ineq3}
\end{align}
Thus, \eqref{eq:lem3ineq3} implies that
\begin{align}
\limsup\limits_{\substack{y_1 \to y_1^* \\t \downarrow 0}}\inf_{d'\to d} &\frac{|h(y_1+td')-h(y_1)|}{t} \nonumber\\
&+ \frac{\sqrt{\tilde{b}^2-(td')^2} - \sqrt{\tilde{b}^2}}{t} \leq 0, \label{eq:lem3ineq4}
\end{align}
where we take $d>0$ small so that the assumption of having a small $d'$ holds. Since the limit of the second term in \eqref{eq:lem3ineq4} is zero, we get
\begin{align}
\limsup\limits_{\substack{y_1 \to y_1^* \\t \downarrow 0}}\inf_{d'\to d} \frac{|h(y_1+td')-h(y_1)|}{t} = 0.
\end{align}
However, this is a contradiction to the fact that either $s^*\in \partial h(y_1^*)$ or $s^*\in {-\partial}{(-h(y_1^*))}$ holds.

\section{Proof of Lemma~\ref{lem:continuum2}}\label{proof:lemmaContinuum2}

In order to prove this result, we assume that there exists a continuum of decoder actions whose support does not extend to the boundaries of the support and then reach a contradiction. Let $\boldsymbol{y}^\alpha=[y_1^\alpha,y_2^\alpha]^T$ and $\boldsymbol{y}^\beta=[y_1^\beta,y_2^\beta]^T$ be decoder actions on the continuum with $y_1^\alpha<y_1^\beta$ and $y_2^\alpha=y_2^\beta$ such that there exist decoder actions for all values of $y_1$ satisfying $y_1^\alpha\leq y_1 \leq y_1^\beta$ and $y_2=y_2^\alpha$ where $x_1^L(y_2^\alpha)<y_1^\alpha$ and $y_1^\beta<x_1^U(y_2^\beta)$. As mentioned earlier, $x_1^L(x_2)$ and $x_1^U(x_2)$ respectively denote the lower and upper boundaries of the support for $X_1$ given that $X_2=x_2$. In addition, suppose that there exists $\delta>0$ such that there is no decoder action with $y_1^\beta<y_1<y_1^\beta+\delta$ and $y_2=y_2^\beta$. Similarly, suppose that there exists $\tilde{\delta}>0$ such that there is no decoder action with $y_1^\alpha-\tilde{\delta}<y_1<y_1^\alpha$ and $y_2=y_2^\alpha$. In the following, we focus on the decoder action $\boldsymbol{y}^\beta$ and show that the assumption of $y_1^\beta<x_1^U(y_2^\beta)$ leads to a contradiction. A similar approach can be taken for the decoder action $\boldsymbol{y}^\alpha$ to prove that the assumption of $x_1^L(y_2^\alpha)<y_1^\alpha$ leads to a contradiction. 

Let $\mathcal{B}^\beta$ denote the bin corresponding to the decoder action $\boldsymbol{y}^\beta$, i.e., $\boldsymbol{y}^\beta=\mathbb{E}[\boldsymbol{X}|\boldsymbol{X}\in\mathcal{B}^\beta ]$. We will obtain a contradiction that $\mathcal{B}^\beta$ contains observations with $x_1>y_1^\beta$, whereas it does not contain any observation with $x_1<y_1^\beta$. This is a contradiction to $\boldsymbol{y}^\beta=\mathbb{E}[\boldsymbol{X}|\boldsymbol{X}\in\mathcal{B}^\beta ]$ since we assume that every non-empty open set has a positive probability measure in Assumption~\ref{assumption}. 

We first show that $\mathcal{B}^\beta$ does not contain any observations with $x_1<y_1^\beta$. Take an observation $\boldsymbol{x}$ with $x_1<y_1^\beta$. Since the continuum of decoder actions is supported on $y_1\in [y_1^\alpha,y_1^\beta]$ and $y_2=y_2^\beta$, we can find a decoder action $\boldsymbol{y}^\nu$ on the continuum with coordinates $y_1^\nu= \mathrm{max}\{x_1,y_1^\alpha\}$ and $y_2^\nu= y_2^\beta$. For this decoder action, when $x_1<y_1^\beta$, we have that $c^e_t(\boldsymbol{x},\boldsymbol{y}^\nu)<c^e_t(\boldsymbol{x},\boldsymbol{y}^\beta)$. This implies that any observation with $x_1<y_1^\beta$ cannot be an element of $\mathcal{B}^\beta$. 

Next, we prove that $\mathcal{B}^\beta$ must contain observations with $x_1>y_1^\beta$ under the assumed configuration, which, however, contradicts $\boldsymbol{y}^\beta=\mathbb{E}[\boldsymbol{X}|\boldsymbol{X}\in\mathcal{B}^\beta ]$. While proving that $\mathcal{B}^\beta$ must contain observations with $x_1>y_1^\beta$, we use the result from Lemma~\ref{lem:geoCon} which imposes the condition in \eqref{eq:geoConTransformed} in the transformed coordinate system considering any two decoder actions at a Nash equilibrium. In particular, for any decoder action $\boldsymbol{y}^\eta$, we need $g(\boldsymbol{y}^\beta,\boldsymbol{y}^\eta)\geq 0$ where $g(\cdot,\cdot)$ is defined in \eqref{eq:geoConTransformed}. Note that other decoder actions impose additional constraints on the region where decoder actions can exist at a Nash equilibrium. Nonetheless, for our purpose, it suffices to use the condition $g(\boldsymbol{y}^\beta,\boldsymbol{y}^\eta)\geq 0$. Our aim is to show that for all $\boldsymbol{y}^\eta$ satisfying $g(\boldsymbol{y}^\beta,\boldsymbol{y}^\eta)\geq 0$, it is possible to find a region of observations with $x_1>y_1^\beta$ where $c^e_t(\boldsymbol{x},\boldsymbol{y}^\beta)<c^e_t(\boldsymbol{x},\boldsymbol{y}^\eta)$. This implies that the intersection of these observation regions must be the bin for $\boldsymbol{y}^\beta$, i.e., $\mathcal{B}^\beta$. To conclude the result, we will show that this intersection is not empty. Note that $c^e_t(\boldsymbol{x},\boldsymbol{y}^\beta)<c^e_t(\boldsymbol{x},\boldsymbol{y}^\eta)$ is equivalent to 
\begin{align}
(y_1^\eta-y_1^\beta)&(y_1^\eta+y_1^\beta-2x_1)\nonumber\\
&+(y_2^\eta-y_2^\beta)(y_2^\eta+y_2^\beta+2\tilde{b}-2x_2) > 0.\label{eq:desiredCond}
\end{align}
We will use this equivalent expression in the remainder of the proof.

When $y_2^\eta= y_2^\beta$, due to the discontinuity assumption at $\boldsymbol{y}^\beta$, for any observation with $y_1^\beta<x_1<y_1^\beta+\delta/2$, the inequality $c^e_t(\boldsymbol{x},\boldsymbol{y}^\beta)<c^e_t(\boldsymbol{x},\boldsymbol{y}^\eta)$ holds. Therefore, it is sufficient to consider decoder actions with $y_2^\eta\neq y_2^\beta$. We can express a decoder action $\boldsymbol{y}^\eta$ satisfying $y_2^\eta\neq y_2^\beta$ and $g(\boldsymbol{y}^\beta,\boldsymbol{y}^\eta)\geq 0$ as $
\boldsymbol{y}^\eta =\boldsymbol{y}^\beta + \lambda (\boldsymbol{y}^\gamma-\boldsymbol{y}^\beta)
$ for some $\lambda\geq 1$ where $\boldsymbol{y}^\gamma$ satisfies $g(\boldsymbol{y}^\beta,\boldsymbol{y}^\gamma)= 0$. In particular, we have
\begin{align*}
&(y_1^\eta-y_1^\beta)^2 + (y_2^\eta-y_2^\beta)^2 - 2\tilde{b}|y_2^\eta-y_2^\beta|\\
&=\lambda^2(y_1^\gamma-y_1^\beta)^2 + \lambda^2(y_2^\gamma-y_2^\beta)^2 - 2|\lambda|\,|y_2^\gamma-y_2^\beta|\tilde{b}\\
&\stackrel{(a)}{=}\lambda^2(2\tilde{b}|y_2^\gamma-y_2^\beta|) - 2|\lambda|\,|y_2^\gamma-y_2^\beta|\tilde{b}\\
&=(2\tilde{b}|y_2^\gamma-y_2^\beta|) (\lambda^2 - |\lambda|)
\end{align*}
where $(a)$ follows from $g(\boldsymbol{y}^\beta,\boldsymbol{y}^\gamma)=0$, and the final expression implies that we can write $\boldsymbol{y}^\eta =\boldsymbol{y}^\beta + \lambda (\boldsymbol{y}^\gamma-\boldsymbol{y}^\beta)$ with $\lambda\geq 1$ so that $g(\boldsymbol{y}^\beta,\boldsymbol{y}^\eta)\geq 0$ holds. It is seen that when $\lambda >1$, for a given observation $\boldsymbol{x}$ satisfying $c^e_t(\boldsymbol{x},\boldsymbol{y}^\beta)<c^e_t(\boldsymbol{x},\boldsymbol{y}^\gamma)$, the following holds: 
\begin{align*}
&(y_1^\eta-y_1^\gamma)(y_1^\eta+y_1^\gamma-2x_1)\\
&+(y_2^\eta-y_2^\gamma)(y_2^\eta+y_2^\gamma+2\tilde{b}-2x_2) \\
&=(\lambda-1)(y_1^\gamma-y_1^\beta)\big((\lambda+1)y_1^\gamma+(1-\lambda)y_1^\beta-2x_1\big)\\
&+(\lambda-1)(y_2^\gamma-y_2^\beta)\big((\lambda+1)y_2^\gamma+(1-\lambda)y_2^\beta+2\tilde{b}-2x_2\big) \\
&>(\lambda-1)(y_1^\gamma-y_1^\beta)\big((\lambda+1)y_1^\gamma+(1-\lambda)y_1^\beta-y_1^\gamma-y_1^\beta\big)\\
&+(\lambda-1)(y_2^\gamma-y_2^\beta)\big((\lambda+1)y_2^\gamma+(1-\lambda)y_2^\beta-y_2^\gamma-y_2^\beta\big) \\
&=\lambda(\lambda-1)\big( (y_1^\gamma-y_1^\beta)^2 +(y_2^\gamma-y_2^\beta)^2\big)>0,
\end{align*}
where the first inequality is due to $c^e_t(\boldsymbol{x},\boldsymbol{y}^\beta)<c^e_t(\boldsymbol{x},\boldsymbol{y}^\gamma)$ and $\lambda>1$, and the last inequality follows from $\boldsymbol{y}^\beta\neq\boldsymbol{y}^\gamma$ and $\lambda>1$. This shows that $c^e_t(\boldsymbol{x},\boldsymbol{y}^\beta)<c^e_t(\boldsymbol{x},\boldsymbol{y}^\gamma)$ implies $c^e_t(\boldsymbol{x},\boldsymbol{y}^\gamma)<c^e_t(\boldsymbol{x},\boldsymbol{y}^\eta)$. Hence, it is sufficient to consider decoder actions $\boldsymbol{y}^\gamma$ with $g(\boldsymbol{y}^\beta,\boldsymbol{y}^\gamma)=0$.  

\begin{figure}
\centering
\includegraphics[width=0.5\linewidth]{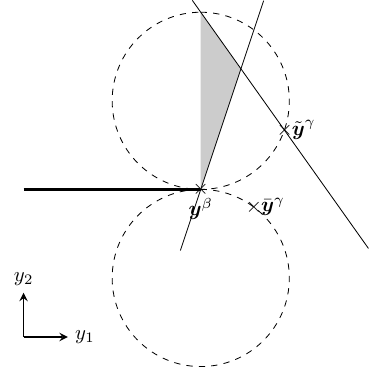}
\caption{Illustration of the proof technique employed in Lemma~\ref{lem:continuum2}. Here, the horizontal solid line represents a continuum of decoder actions, and the decoder action on this continuum with the largest $y_1$ coordinate is denoted by $\boldsymbol{y}^\beta$. Inside the dashed circles, there cannot be a decoder action due to the condition imposed by Lemma~\ref{lem:geoCon} that for any $\boldsymbol{y}$, we have $g(\boldsymbol{y}^\beta,\boldsymbol{y})\geq 0$ where $g(\cdot,\cdot)$ is defined in \eqref{eq:geoConTransformed}. If we place decoder actions $\bar{\boldsymbol{y}}^\gamma$ and $\tilde{\boldsymbol{y}}^\gamma$ on the dashed circles, then the shaded area must be the bin for $\boldsymbol{y}^\beta$, i.e., $\mathcal{B}^\beta$. However, the centroid of this shaded area cannot be $\boldsymbol{y}^\beta$, which is a contradiction.}  
\label{fig:continuum}
\end{figure}

Next, we show that for any decoder action $\boldsymbol{y}^\gamma$ satisfying $g(\boldsymbol{y}^\beta,\boldsymbol{y}^\gamma)= 0$, there exists a nonempty region of observations with $c^e_t(\boldsymbol{x},\boldsymbol{y}^\beta)<c^e_t(\boldsymbol{x},\boldsymbol{y}^\gamma)$ and $x_1>y_1^\beta$. Towards that goal, we consider different cases and treat each of these cases separately. In the following, we take $y_2^\beta<x_2<y_2^\beta+2\tilde{b}$ and specify a nontrivial interval for $x_1$ so that the resulting observation satisfies the desired property. 

\begin{enumerate}[(i)]
\item Let $\boldsymbol{y}^\gamma$ satisfy $y_2^\beta<y_2^\gamma<y_2^\beta+2\tilde{b}$ and $y_1^\gamma= y_1^\beta + ((y_2^\gamma-y_2^\beta)(y_2^\beta-y_2^\gamma+2\tilde{b}))^{1/2}$. Let
\begin{align}
y_1^\beta< x_1 <y_1^\beta+ \frac{(y_2^\gamma-y_2^\beta)(y_2^\beta-x_2+2\tilde{b})}{(y_1^\gamma-y_1^\beta)}.\label{eq:x1Case1}
\end{align}
After some manipulations, it can be shown that \eqref{eq:desiredCond} and equivalently $c^e_t(\boldsymbol{x},\boldsymbol{y}^\beta)<c^e_t(\boldsymbol{x},\boldsymbol{y}^\gamma)$ hold for any observation that satisfies \eqref{eq:x1Case1} and $y_2^\beta<x_2<y_2^\beta+2\tilde{b}$. Notice also that if \eqref{eq:x1Case1} holds for some $y_2^\gamma$, then it automatically holds for any $\tilde{y}_2^\gamma$ satisfying $y_2^\gamma <\tilde{y}_2^\gamma<y_2^\beta+2\tilde{b}$. Therefore, multiple decoder actions satisfy the assumptions of this case, it is sufficient to consider the one with the minimum $y_2$ coordinate. 

\item Let $\boldsymbol{y}^\gamma$ be such that $y_2^\beta< y_2^\gamma \leq y_2^\beta+2\tilde{b}$ and $y_1^\gamma= y_1^\beta - ((y_2^\gamma-y_2^\beta)(y_2^\beta-y_2^\gamma+2\tilde{b}))^{1/2}$ hold. In this case, one can show that for any observation with $y_1^\beta< x_1$ and $y_2^\beta<x_2<y_2^\beta+2\tilde{b}$, the inequality $c^e_t(\boldsymbol{x},\boldsymbol{y}^\beta)<c^e_t(\boldsymbol{x},\boldsymbol{y}^\gamma)$ holds. 
\item Let $\boldsymbol{y}^\gamma$ satisfy $y_2^\beta-2\tilde{b}< y_2^\gamma<y_2^\beta$ and $y_1^\gamma= y_1^\beta +((y_2^\gamma-y_2^\beta)(y_2^\beta-y_2^\gamma-2\tilde{b}))^{1/2}$. Let 
\begin{align}
y_1^\beta<x_1  <  y_1^\beta+\frac{(y_2^\beta-y_2^\gamma)(x_2-y_2^\beta)}{(y_1^\gamma-y_1^\beta)}.\label{eq:x1Case3}
\end{align}
After some manipulations, it can be shown that \eqref{eq:desiredCond} and equivalently $c^e_t(\boldsymbol{x},\boldsymbol{y}^\beta)<c^e_t(\boldsymbol{x},\boldsymbol{y}^\gamma)$ hold for any observation that satisfies \eqref{eq:x1Case3} and $y_2^\beta<x_2<y_2^\beta+2\tilde{b}$. In addition, notice that if \eqref{eq:x1Case3} holds for some $y_2^\gamma$, then it automatically holds for any $\tilde{y}_2^\gamma$ satisfying $y_2^\gamma-2\tilde{b} <\tilde{y}_2^\gamma<y_2^\gamma$. Therefore, multiple decoder actions satisfy the assumptions of this case, it is sufficient to consider the one with the maximum $y_2$ coordinate.
\item Let $\boldsymbol{y}^\gamma$ be such that $y_2^\beta-2\tilde{b}\leq y_2^\gamma<y_2^\beta$ and $y_1^\gamma= y_1^\beta -((y_2^\gamma-y_2^\beta)(y_2^\beta-y_2^\gamma-2\tilde{b}))^{1/2}$. In this case, one can show that for any observation with $y_1^\beta< x_1$ and $y_2^\beta<x_2<y_2^\beta+2\tilde{b}$, the inequality $c^e_t(\boldsymbol{x},\boldsymbol{y}^\beta)<c^e_t(\boldsymbol{x},\boldsymbol{y}^\gamma)$ holds. 
\end{enumerate}

As a result, for any decoder action $\boldsymbol{y}^\eta$ satisfying $g(\boldsymbol{y}^\beta,\boldsymbol{y}^\eta)\geq 0$ and for a given $x_2$ with $y_2^\beta<x_2<y_2^\beta+2\tilde{b}$, a nontrivial interval exists for $x_1$ with $x_1>y_1^\beta$ such that $c^e_t(\boldsymbol{x},\boldsymbol{y}^\beta)<c^e_t(\boldsymbol{x},\boldsymbol{y}^\eta)$ holds. This implies that there exists a nonempty region of observations with $x_1>y_1^\beta$ that must belong to $\mathcal{B}^\beta$. Fig.~\ref{fig:continuum} illustrates this region of observations for an example scenario. Note that as long as $y_1^\beta<x_1^U(y_2^\beta)$, a nonempty subset of this observation region is in the support of the joint distribution. As a result, we obtain a contradiction to $\boldsymbol{y}^\beta=\mathbb{E}[\boldsymbol{X}|\boldsymbol{X}\in\mathcal{B}^\beta ]$ since $\mathcal{B}^\beta$ does not contain any observation with $x_1<y_1^\beta$, and $\mathcal{B}^\beta$ contains a nonempty region of observations with $x_1>y_1^\beta$. 

\section{Proof of Theorem~\ref{thm:main}}\label{proof:main} 

Suppose without loss of generality that $\mathbb{E}[M_1]=0$. In the proof, we consider the equivalent formulation in Lemma~\ref{lem:equivalent}.

\begin{enumerate}[(i)]
\item Since $M_1$ and $M_2$ are independent random variables, the problem decouples into two one-dimensional cheap talk problems where one of them involves an encoder with a zero bias. Hence, revealing the source corresponding to a zero bias leads to a linear Nash equilibrium.
\item If the source distribution is Gaussian, $X_1$ and $X_2$ are independent random variables. Thus, we obtain decoupled one-dimensional cheap talk problems where one of them involves an encoder with a zero bias. Then, revealing the random variable corresponding to a zero bias (i.e., $X_1$) yields an informative Nash equilibrium where the encoder is linear.

Now, suppose that the source distribution is not Gaussian. In this case, we show that there does not exist a Nash equilibrium with an encoding policy $z=\gamma^e(\boldsymbol{m})=\alpha_1m_2-\alpha_2 m_1$ for any scalars $\alpha_1$ and $\alpha_2$. From Lemma~\ref{lem:continuum1} and Lemma~\ref{lem:continuum2}, we know that a continuum of actions must have a constant $y_2$ coordinate, say $\kappa$, and must be supported for all values of $y_1$ in the support of $X_1$ given that $X_2=\kappa$. This implies that a necessary condition for a Nash equilibrium with an encoding policy $z=\gamma^e(\boldsymbol{m})=\alpha_1m_2-\alpha_2 m_1$ is given by $\mathbb{E}[X_2|\bar{X}=\bar{x}]=\mathbb{E}[b_1M_1+b_2M_2|\alpha_1M_2-\alpha_2 M_1=\bar{x}] = 0$ for all $\bar{x}$ where $\bar{X}\triangleq \alpha_1M_2-\alpha_2 M_1$. If $\alpha_1=0$, $\alpha_2=0$, or $\frac{\alpha_1}{\alpha_2}\neq \frac{b_1}{b_2}$, one can decompose $X_2$ to show that the condition of $\mathbb{E}[X_2|\bar{X}=\bar{x}]=0$ for all $\bar{x}$ is always violated. It follows that there does not exist a Nash equilibrium with a linear encoding policy $z=\gamma^e(\boldsymbol{m})=\alpha_1m_2-\alpha_2 m_1$ when $\alpha_1=0$, $\alpha_2=0$, or $\frac{\alpha_1}{\alpha_2}\neq \frac{b_1}{b_2}$. It remains to investigate conditions under which an encoding policy $z=\gamma^e(\boldsymbol{m})=b_1m_2-b_2 m_1$ leads to a Nash equilibrium.

If $b_1\neq 0$, $b_2\neq 0$ and $|b_1|\neq |b_2|$, the condition of $\mathbb{E}[b_1M_1+b_2M_2|b_1M_2-b_2M_1=x_1]=\mathbb{E}[X_2|X_1=x_1]=0$ for all $x_1$ requires that the source distribution is Gaussian \citep[Theorem~5.3.1]{KaganBook}. Hence, if the encoder reveals $X_1=b_1M_2-b_2M_1$ completely without giving additional information, we obtain a single continuum that contains decoder actions with different second coordinates. Since this contradicts with Lemma~\ref{lem:continuum1}, there cannot be a Nash equilibrium where the encoder conveys $X_1=b_1M_2-b_2M_1$ only. In addition, having more than one continuum of decoder actions, each with a constant second coordinate, implies that $\mathbb{E}[X_2|X_1=x_1]=0$ for all $x_1$. Hence, it is not possible to have a Nash equilibrium with more than one continuum of decoder actions.

\item In the case of $b_1=b_2$, the condition of $\mathbb{E}[X_2|X_1=x_1]=0$ for all $x_1$ requires that the source distribution is symmetric (almost everywhere) \citep[Theorem~5.3.1]{KaganBook}. It follows that if the source distribution is not symmetric, there does not exist an informative linear Nash equilibrium. Now, suppose that the source distribution is symmetric. Let the encoding policy be given by $z=\tilde{\gamma}^e(\boldsymbol{x})=x_1$. In other words, the encoder reveals $X_1$ completely without giving any additional information. Then, the best response of the decoder yields a single continuum of decoder actions. In particular, there only exist decoder actions for all values of $y_1\in [x_1^L(0),x_1^U(0)]$ and $y_2=0$ where $x_1^L(0)$ and $x_1^U(0)$ respectively denote lower and upper boundaries of the support for $X_1$ given that $X_2=0$. Now, we suppose that there only exist decoder actions for all values of $y_1\in [x_1^L(0),x_1^U(0)]$ and $y_2=0$, and we wish to obtain the best response of the encoder to these decoder actions. For any given decoder actions $\tilde{\boldsymbol{y}}$ and $ \boldsymbol{\bar{y}}$ satisfying $\tilde{\boldsymbol{y}}\neq \boldsymbol{\bar{y}}$, it must be that $\tilde{y}_1\neq \bar{y}_1$ as every decoder action is assumed to have $y_2=0$. Note also that for any given observation $\boldsymbol{x}$ in the support of the joint distribution, there exists a decoder action $\boldsymbol{y}=[y_1,y_2]^T$ with $y_1=x_1$ and $y_2=0$. These imply that if a given observation $\boldsymbol{x}$ satisfies $x_1=\tilde{y}_1$, it follows that
\begin{align*}
c_t^e(\boldsymbol{x},\tilde{\boldsymbol{y}})&=(x_1-\tilde{y}_1)^2 + (x_2-\tilde{y}_2-\tilde{b})^2 \\
&< (x_1-\bar{y}_1)^2 + (x_2-\bar{y}_2-\tilde{b})^2=c_t^e(\boldsymbol{x},\bar{\boldsymbol{y}})
\end{align*}
for any decoder actions $\bar{\boldsymbol{y}}$ and $\tilde{\boldsymbol{y}}$ satisfying $\bar{\boldsymbol{y}}\neq \tilde{\boldsymbol{y}}$. Therefore, if we denote the bin corresponding to a decoder action $\boldsymbol{y}$ by $B_{\boldsymbol{y}}$, we get $B_{\boldsymbol{y}} = \{ \boldsymbol{x} \,|\, x_1=y_1\text{ and } x_2\in\mathbb{R} \}$. This means that the best response of the encoder is to reveal the value of $X_1$ completely without giving any additional information. Hence, the encoding policy $z=\tilde{\gamma}^e(\boldsymbol{x})=x_1$ and the decoding policy $\boldsymbol{y}=\tilde{\gamma}^d(z)=[z,0]^T$ form a Nash equilibrium as they are best response maps of each other. 

\item From \citep[Theorem~5.3.1]{KaganBook}, we know that $\mathbb{E}[X_2|X_1=x_1]=0$ for all $x_1$ regardless of the source distribution when $b_1=-b_2$. In this case, a similar analysis can be carried out to show that there always exists an informative linear Nash equilibrium for any source distribution when $b_1=-b_2$.
\end{enumerate}

\section{Proof of Theorem~\ref{thm:gaussian}}\label{proof:gaussian} 

By applying a suitable linear transformation of variables, one can obtain an equivalent problem. In this equivalent problem, the aim is to convey a sequence of independent Gaussian sources, and the bias is zero for all sources except one. Due to the independence of these sources, the problem decouples. By using the result of \citep[Theorem~4]{NumberOfBinsArxiv}, it follows that there exists a Nash equilibrium where the encoder uses a quantization policy with any number of bins for the source corresponding to a non-zero bias. On the other hand, since the bias is zero for all other sources, revealing these sources leads to a Nash equilibrium. 

While it is possible to apply a transformation of variables for any $n\geq 2$, we specify a transformation of variables for $n=3$ dimensional scenario as an example. In particular, consider $\boldsymbol{X} = \mathcal{T}\boldsymbol{M}$ and  $\boldsymbol{U}=\mathcal{T}^{-1}\boldsymbol{Y}$ where 
\begin{align*}
&\mathcal{T} \triangleq \frac{1}{\sqrt{b_1^2+b_2^2+b_3^2}}\\
&\times \begin{bmatrix}
\frac{b_2\sqrt{b_1^2+b_2^2+b_3^2}}{\sqrt{b_1^2+b_2^2}} & \frac{-b_1\sqrt{b_1^2+b_2^2+b_3^2}}{\sqrt{b_1^2+b_2^2}} & 0\\
\frac{b_1b_3}{{\sqrt{b_1^2+b_2^2}}} & \frac{b_2b_3}{{\sqrt{b_1^2+b_2^2}}} & 
-\sqrt{b_1^2+b_2^2}
\\
b_1 & b_2 & b_3
\end{bmatrix}=(\mathcal{T}^{-1})^T.
\end{align*}
Under this transformation of variables, the objective function of the encoder becomes
\begin{align*}
c^e(\boldsymbol{m},\boldsymbol{u}) = &
(x_1-y_1)^2
+ (x_2-y_2)^2\\
&+(x_3-y_3-(b_1^2+b_2^2+b_3^2)^{1/2})^2 \triangleq c^e_t(\boldsymbol{x},\boldsymbol{y}).
\end{align*}
For the objective function of the decoder, we get $
c^d(\boldsymbol{m},\boldsymbol{u}) = \lVert \boldsymbol{m}-\boldsymbol{u}\rVert^2=\lVert \boldsymbol{x}-\boldsymbol{y}\rVert^2\triangleq c^d_t(\boldsymbol{x},\boldsymbol{y})
$. Since $X_1,X_2,X_3$ are independent random variables, we obtain decoupled one-dimensional cheap talk problems where the biases for $X_1$ and $X_2$ are zero, and the bias for $X_3$ is non-zero. Thus, revealing $X_1$ and/or $X_2$ and applying a signaling game policy for $X_3$ yield a Nash equilibrium.

\section{Proof of Theorem~\ref{thm:gaussiancorrelated}}\label{proof:gaussiancorrelated} 

If the condition in the statement of the theorem holds, the sources $X_1=b_1M_2-b_2M_1$ and $X_2=b_1M_1+b_2M_2$ in the equivalent problem become independent. Therefore, in the equivalent problem, we get decoupled one-dimensional cheap talk problems where there is a zero bias for $X_1$ and a non-zero bias for $X_2$. Thus, revealing $X_1$ yields an informative linear Nash equilibrium. 

\section{Proof of Theorem~\ref{thm:sym}}\label{proof:sym}

If the encoder uses $M_1,\dots,M_{\tilde{n}}$ with $\tilde{n}<n$ in constructing its linear policy as described below and gives no information related to $M_{\tilde{n}+1},\dots,M_n$, we obtain a linear Nash equilibrium where the encoder reveals $(\tilde{n}-1)$ dimensions completely. Suppose without loss of generality that $\mathbb{E}[M_1]=0$. We first apply a linear transformation of variables. In a similar manner to the two-dimensional case, we obtain an equivalent problem in a transformed coordinate system. In particular, let
\begin{align}
\mathcal{T} \triangleq  
\begin{bmatrix}
\frac{1}{\sqrt{2}} & \frac{-1}{\sqrt{2}} & 0 & \dots & \\
\frac{1}{\sqrt{2\times 3}} & \frac{1}{\sqrt{2\times 3}} & \frac{-2}{\sqrt{2\times 3}} & 0 & \dots \\
 &  & \vdots & \\
\frac{1}{\sqrt{(n-1)\times n}} &  & \dots & \frac{1}{\sqrt{(n-1)\times n}} & \frac{-(n-1)}{\sqrt{(n-1)\times n}} \\
\frac{1}{\sqrt{n}} & & \dots & & \frac{1}{\sqrt{n}}
\end{bmatrix}.\label{eq:transformNDim}
\end{align}
Since $\mathcal{T}^{-1}=\mathcal{T}^T$ and $\mathcal{T}\boldsymbol{b}=[0,\dots,0,\sqrt{n}b]^T$, the cost function of the encoder for the equivalent problem becomes
\begin{align}
c^e&(\boldsymbol{m},\boldsymbol{u})
=
\sum_{k=1}^n (m_k-u_k-b)^2 \nonumber \\
&= 
\sum_{k=1}^{n-1} (x_k-y_k)^2 + (x_n-y_n-\sqrt{n}b)^2 \triangleq c^e_t(\boldsymbol{x},\boldsymbol{y}).\nonumber
\end{align}
The cost function of the decoder in this transformed coordinate system is given by $c^d(\boldsymbol{m},\boldsymbol{u}) =\lVert \boldsymbol{m}-\boldsymbol{u}\rVert^2 = \lVert \boldsymbol{x}-\boldsymbol{y}\rVert^2 \triangleq c_t^d(\boldsymbol{x},\boldsymbol{y})$. It is seen that there is no bias for $X_1,\dots,X_{n-1}$, and there is a non-zero bias for $X_n$ considering the cost function of the encoder in this transformed coordinate system. 

We note that revealing $X_1,\dots,X_{n-1}$ is equivalent to revealing $M_1-M_2$, $M_2-M_3$, $\dots$, $M_{n-1}-M_n$. In the following, we show that if the encoder reveals these random variables, the optimal estimate for $X_n$ at the decoder becomes zero. Towards that goal, let $\tilde{M}_1 \triangleq -M_n$, $\tilde{M}_2 \triangleq -M_{n-1}$, $\dots$, $\tilde{M}_n \triangleq -M_1$ and observe that $\tilde{M}_1,\dots,\tilde{M}_n$ have the same distribution as $M_1,\dots,M_n$ due to the symmetry of the source distribution. Hence, we get
\begin{align*}
&\mathbb{E}[M_1+\dots+M_n | M_1-M_2,\dots,M_{n-1}-M_n]\\
&=-\mathbb{E}[\tilde{M}_n+\dots+\tilde{M}_1 |\tilde{M}_{n-1}-\tilde{M}_n,\dots,\tilde{M}_{1}-\tilde{M}_2 ].
\end{align*}
This proves that  the conditional mean of $X_n$ given that $X_1,\dots,X_{n-1}$ are revealed is zero. We know that in the transformed coordinate system, there is a non-zero bias only for $X_n$. If the encoder reveals $X_1,\dots,X_{n-1}$ completely, the best response of the decoder is to use these revealed parameters as the corresponding estimates, and the action taken for $X_n$ is zero. In other words, if we have $\boldsymbol{z}=\tilde{\gamma}^e(\boldsymbol{x}) = [x_1,\dots,x_{n-1}]^T$ as the encoding policy, the best response of the decoder becomes $\boldsymbol{y}=\tilde{\gamma}^d(\boldsymbol{z}) = [z_1,\dots,z_{n-1},0]^T$. Since we consider the Nash equilibrium concept, it is required to take the best response of the encoder into account, as well. Towards that goal, one can construct quantization bins as in the proof of Theorem~\ref{thm:main} to prove the result. In fact, it is also possible to see the result by only looking at the policies. In particular, suppose that the decoder uses the policy $\boldsymbol{y}=\tilde{\gamma}^d(\boldsymbol{z}) = [z_1,\dots,z_{n-1},0]^T$. This means that the decoder uses $y_1,\dots,y_{n-1}$ as its estimates for $X_1,\dots,X_{n-1}$, and the estimate for $X_n$ is zero regardless of the transmitted message. Since the encoder wishes accurate estimations of $X_1,\dots,X_{n-1}$ at the decoder without any bias, the best response of the encoder to the decoding policy $\boldsymbol{y}=\tilde{\gamma}^d(\boldsymbol{z})=[z_1,\dots,z_{n-1},0]^T$ becomes $\boldsymbol{z}=\tilde{\gamma}^e(\boldsymbol{x})=[x_1,\dots,x_{n-1}]^T$. Although the encoder has a bias regarding $X_n$, the encoder cannot affect the corresponding estimate at the decoder since the decoder action $y_n$ is zero regardless of the encoded message. This implies that the encoding policy $\boldsymbol{z}=\tilde{\gamma}^e(\boldsymbol{x})=[x_1,\dots,x_{n-1}]^T$ and the decoding policy $\boldsymbol{y}=\tilde{\gamma}^d(\boldsymbol{z})=[z_1,\dots,z_{n-1},0]^T$ are best response maps of each other. Hence, an encoding policy that completely reveals $X_1,\dots,X_{n-1}$ leads to a Nash equilibrium.

\section{Proof of Theorem~\ref{thm:infoTheoretic}}\label{proof:infoTheoretic}

Suppose without loss of generality that $\mathbb{E}[M_1]=0$. Consider the transformation of variables in \eqref{eq:transformNDim}. We obtain an equivalent problem as in Lemma~\ref{lem:equivalent} where the linear transformation $\mathcal{T}$ is fixed as in \eqref{eq:transformNDim}, and the encoder and decoder design $\tilde{\gamma}^e(\cdot)$ and $\tilde{\gamma}^d(\cdot)$, respectively. The random variables $X_1,\dots,X_n$ defined by \eqref{eq:transformNDim} are i.i.d. and follow the same distribution as $M_1,\dots,M_n$. Thus, the problem decouples to $n$ one-dimensional cheap talk problems where there is a non-zero bias only for one of the problems. Suppose that the encoder does not reveal information related to $X_n$, which is the source corresponding to a non-zero bias. In this case, the rate of the original problem is identical to that of a team theoretic problem with $(n-1)$ i.i.d. sources. Since the term that contributes to the objective of the encoder is $\mathbb{E}[(X_n-Y_n-\sqrt{n}b)^2]=nb^2$ in the case that no information is conveyed related to $X_n$, we obtain an additional $b^2$ term for the encoder's distortion bound. As a result, we get
\begin{align*}
\frac{\mathbb{E}[\sum_{i=1}^n(M_i-U_i-b)^2]}{n} = \frac{\mathbb{E}[\sum_{i=1}^{n-1}(X_i-Y_i)^2]}{n} +b^2.
\end{align*}
By taking the limit of both sides, we have
\begin{align*}
&\lim_{n\to\infty} \frac{\mathbb{E}[\sum_{i=1}^n(M_i-U_i-b)^2]}{n} \\
&=\lim_{n\to\infty}  \frac{\mathbb{E}[\sum_{i=1}^{n-1}(X_i-Y_i)^2]}{n}+b^2\\
&=\lim_{n\to\infty}  \frac{\mathbb{E}[\sum_{i=1}^{n-1}(X_i-Y_i)^2]}{n-1} + b^2,
\end{align*}
where the last equality follows since $n\to\infty$. Hence, we obtain a reduced team theoretic problem where the encoder wishes to convey an i.i.d. source $X_1,\dots,X_{n-1}$ with a zero bias. Therefore, if the pair $(R_T,D_T)$ is achievable in a team theoretic setup, $R=R_T$ and $D_e\geq D_T+b^2$ are achievable for the original game theoretic setup. For the distortion bound of the decoder, we use the relation $\mathbb{E}[\sum_{i=1}^n(M_i-U_i-b)^2]=\mathbb{E}[\sum_{i=1}^n(M_i-U_i)^2]+b^2n$.

\section{Proof of Theorem~\ref{thm:RDGame}}\label{proof:RDGame} 
If we take $D_T=\min\{D_e-b^2,D_d\}$ for the team theoretic setup in Theorem~\ref{thm:infoTheoretic}, the corresponding rate $R_T$ is achievable for the game theoretic setup. Thus, we get 
\begin{align*}
&\inf\{R \,|\, (R,D_e,D_d) \text{ is achievable}\}\\
&\leq \inf\{R \,|\, (R,\min\{D_e-b^2,D_d\}) \text{ is achievable}\},
\end{align*}
which leads to the bound in the statement of the theorem via Lemma~\ref{thm:RDTeam}.

\end{appendices}

\bibliographystyle{dcu}

{\footnotesize\bibliography{MultiCheapJrnl}}



\end{document}